\documentclass[12pt]{article}
\usepackage{amssymb,amsmath,latexsym}
\usepackage{graphicx}
\usepackage[final]{pdfpages}
\usepackage{verbatim}
\usepackage{color}
\usepackage{amsfonts,amsthm,hyperref,mathrsfs,ulem,tikz}

\usetikzlibrary{arrows,patterns}

\title{Random perturbations of hyperbolic dynamics}

\author{Florian Dorsch, Hermann Schulz-Baldes
\\
\\
{\small Department Mathematik, Friedrich-Alexander-Universit\"at Erlangen-N\"urnberg, Germany}
}

\date{ }

\newtheorem{theo}{Theorem}

\newtheorem{proposi}[theo]{Proposition}
\newtheorem{lemma}[theo]{Lemma}
\newtheorem{coro}[theo]{Corollary}

\numberwithin{figure}{section}

\newcommand{\RM}{{\mathbb R}}
\newcommand{\SM}{{\mathbb S}}

\newcommand{\Tt}{{\cal T}}
\newcommand{\Rr}{{\cal R}}

\newcommand{\one}{{\bf 1}}

\newcommand{\bsm}{\left(\begin{smallmatrix}} 
\newcommand{\esm}{\end{smallmatrix}\right)}  
\definecolor{GR}{rgb}{.35,.7,.35}

\begin{document}

\maketitle

\begin{abstract}
A  sequence of large invertible matrices given by a small random perturbation around a fixed diagonal and positive matrix induces a random dynamics on a high-dimensional sphere. For a certain class of rotationally invariant random perturbations it is shown that the dynamics approaches the stable fixed points of the unperturbed matrix up to errors even if the strength of the perturbation is large compared to the relative increase of nearby diagonal entries of the unperturbed matrix specifying the local hyperbolicity.

\vspace{.1cm}

\end{abstract}


\vspace{.8cm}


\section{Model, main results and comments}

Let us consider the random dynamics on the $\mathsf{L}$-dimensional sphere $\mathbb{S}^{\mathsf{L}}$, $\mathsf{L}\geq 2$,  given by
\begin{equation}
\label{dynamics_sphere}
v_n
\;=\;
\mathcal{T}_n\cdot v_{n-1}\,,\qquad\, n\in\mathbb{N}\,,
\end{equation}
where the action $\cdot\textnormal{ }:\textnormal{GL}(\mathsf{L}+1,\mathbb{R})\times\mathbb{S}^{\mathsf{L}}\rightarrow\mathbb{S}^{\mathsf{L}}$ of the general linear group is
\begin{equation}
\label{eq-ActionDef}
\mathcal{T}\cdot v\;=\; \frac{\mathcal{T}v}{\|\mathcal{T}v\|}
\;,
\end{equation}
and the random matrices $\mathcal{T}_n$ are of the form
\begin{align}
\label{random_matrix}
\mathcal{T}_n
\;=\; 
\mathcal{R}\left(\mathbf{1}\,+\,\lambda r_nU_n\right)\,,\qquad n\in\mathbb{N}
\;.
\end{align}
Here $\mathcal{R}=\textnormal{diag}( \kappa_{\mathsf{L}+1},\dots,\kappa_1)$ is a fixed \textit{unperturbed} positive diagonal matrix  whose entries satisfy $\kappa_1\geq \dots\geq \kappa_{\mathsf{L}+1}>0$ and a \textit{random perturbation} $\lambda r_nU_n$ is given by a coupling constant $\lambda\in[0,1)$, a scalar \textit{radial randomness} $r_n$ and an \textit{angular randomness} induced by orthogonal matrices $U_n$.  The main assumption is that both the $U_n$ and $r_n$ are independent and identically distributed (i.i.d.)  with $\textnormal{O}(\mathsf{L}+1)$-valued and Haar distributed $U_n$ and $[0,1]$-valued $r_n\not\equiv 0$. This includes, \textit{e.g.}, the case $r_n\equiv 1$. Hence the object of study is a particular Markov process on the continuous state space $\mathbb{S}^{\mathsf{L}}$.

\vspace{.2cm}

The above is the standard set-up of the theory of products of random matrices \cite{BL} except that usually the action is studied on  the projective space and not its double cover by $\SM^{\mathsf{L}}$, but for sake of simplicity we suppress this difference. By Furstenberg's Theorem the random action has a unique invariant probability measure $\mu_{r,\lambda}$ on $\mathbb{S}^{\mathsf{L}}$ if $\lambda\not=0$ (see~\cite{BL}, Part A, Theorem III.4.3). This paper is about obtaining further quantitative information about this invariant measure in the special case described above. Hence the paper is thematically located at the interface between random matrix theory, the theory of products of random matrices and random dynamical systems. One of the key technical elements in the proofs is a stochastic order underlying the process \eqref{dynamics_sphere} with $\mathcal{R}=\one$, see Proposition~\ref{prop-StochOd} below.

\vspace{.2cm}

Let us begin by describing the dynamics \eqref{dynamics_sphere} heuristically. The unperturbed deterministic dynamics $\Rr\cdot$ induced by $\Rr$ is maximally hyperbolic if the deterministic local expansion rates
$$
\delta\mathcal{R}_i
\;=\;
\frac{\kappa_{i}-\kappa_{i+1}}{\kappa_{i+1}}
$$ 
are strictly positive for all $i=1,\ldots,\mathsf{L}$. Then there is a simple stable fixed point given by the unit vector $e_{\mathsf{L}+1}$ corresponding to the last component (the fixed point is unique only on projective space).  The deterministic dynamics $\Rr^N \cdot v_0$ converges to the unit vector $e_j$ if $j$ is the largest index such that the $j$th component of the initial condition $v_0$ does not vanish. However, $e_j$ is an unstable fixed point of $\Rr\cdot$ if $j\leq \mathsf{L}$.  All these facts are elementary to check. In the following, we also speak of the unit eigenvector  $e_{\mathsf{L}+2-j}$ of the eigenvalue $\kappa_j$ as the $j$th channel specified by the unperturbed dynamics.  We will {\sl not} assume maximal hyperbolicity in the following. 

\vspace{.2cm}

If now the strength of the perturbation is non-zero and satisfies $\lambda< 2^{-4}\min\{\delta\mathcal{R}_1,\mbox{\small $\frac{1}{2}$}\}$, one can prove that the random dynamics leaves any unstable fixed point and is driven to the vicinity of the stable fixed point in which it then remains. Thus in this case the Furstenberg invariant measure $\mu_{r,\lambda}$ is supported only by a strict subset of $\mathbb{S}^{\mathsf{L}}$, which is a neighborhood of the stable fixed point. More generally, the theorem below states that if $\lambda< 2^{-4}\min\{\delta\mathcal{R}_i,\mbox{\small $\frac{1}{2}$}\}$ for some $i$, then $\textnormal{supp}(\mu_{r,\lambda})$ is a strict subset of $\mathbb{S}^{\mathsf{L}}$. From the proof one can infer that the support is a small (in a quantitative manner) neighborhood of $\{0\}^{\mathsf{L}+1-i}\times\mathbb{S}^{i-1}$.  The main interest of this paper is, however, to analyze the situation where several of the $\delta\mathcal{R}_i$ vanish or are at least all smaller than $\lambda$. Hence the unperturbed dynamics may be merely partially hyperbolic. In this situation the random perturbation is {\sl not} small compared to the local hyperbolicity of $\Rr$. Intuitively, it is clear that the random dynamics may then visit all points on $\mathbb{S}^{\mathsf{L}}$ because the randomness can overcome the hyperbolic character of $\Rr$ and lead to significant escapes from anywhere. This just means that the support of the invariant measure is the whole sphere $\mathbb{S}^{\mathsf{L}}$. This last fact is precisely part of the following first result.

\begin{theo}\label{ac_distribution}
Suppose that $\lambda\in (0,1)$, that the i.i.d. $r_n\not\equiv 0$ are $[0,1]$-valued and that the i.i.d. $U_n$ are Haar distributed on $\textnormal{O}(\mathsf{L}+1)$. Then the Furstenberg measure $\mu_{r,\lambda}$ is absolutely continuous w.r.t. the normalized surface measure $\nu_{\mathsf{L}}$. If $\mathbb{P}(r=0)=0$ holds, then the random variables $v_N\in \mathbb{S}^{\mathsf{L}}$ are distributed absolutely continuously w.r.t. $\nu_{\mathsf{L}}$ on $\mathbb{S}^{\mathsf{L}}$ even for any $N\geq 1$ and initial condition $v_0$. Provided that $\lambda< 2^{-4}\min\{\delta\mathcal{R}_i,\mbox{\small $\frac{1}{2}$}\}$ for some $i=1,\ldots,\mathsf{L}$, the support of $\mu_{r,\lambda}$ is a strict subset of $\mathbb{S}^{\mathsf{L}}$. If $\lambda> \delta\mathcal{R}_i$ for all $i=1,\ldots,\mathsf{L}$ and $1\in\mbox{\rm supp}(r)$, then the support of $\mu_{r,\lambda}$ is the whole sphere $\mathbb{S}^{\mathsf{L}}$. 
\end{theo}

Now let us suppose that the randomness, while being large compared to the local expansion rates $\lambda>\delta\mathcal{R}_i$, is small compared to the expansion rates  
$$
\delta\mathcal{R}_{i,j}
\;=\;
\frac{\kappa_{i}-\kappa_{j}}{\kappa_{j}}
\;,
$$
from channel $i$ to channel $j$ for some $j>i$. Then if $\lambda< \delta\mathcal{R}_{i,j}$, there is some contraction hyperbolicity on this larger scale, even though the local hyperbolicity is dominated by the randomness. Hence a finer analysis of the interplay between the randomness and the hyperbolic unperturbed dynamics is needed. Intuitively, one certainly expects the random dynamics to spend little time in the channel $j$  and this should lead to a small weight of the Furstenberg measure on this channel. Roughly this is what we actually prove below. To state our main result more precisely, we need some further notations. Let us partition the channels into three parts $(\mathsf{L}_{\mathfrak{a}},\mathsf{L}_{\mathfrak{b}},\mathsf{L}_{\mathfrak{c}})\in \mathbb{N}\times\mathbb{N}\times\mathbb{N}$, namely $\mathsf{L}_{\mathfrak{a}}+\mathsf{L}_{\mathfrak{b}}+\mathsf{L}_{\mathfrak{c}}=\mathsf{L}+1$. Each vector $v=(v_1,\dots,v_{\mathsf{L}+1})^{\intercal}\in\mathbb{R}^{\mathsf{L}+1}$ is split into its upper part $\mathfrak{a}(v)\in\mathbb{R}^{\mathsf{L}_{\mathfrak{a}}}$, middle part $\mathfrak{b}(v)\in\mathbb{R}^{\mathsf{L}_{\mathfrak{b}}}$ and lower part $\mathfrak{c}(v)\in\mathbb{R}^{\mathsf{L}_{\mathfrak{c}}}$ via
$$
\mathfrak{a}(v)\;=\; (v_1,\dots, v_{\mathsf{L}_{\mathfrak{a}}})^{\intercal}\;,
\qquad
\mathfrak{b}(v)\;=\; 
(v_{\mathsf{L}_{\mathfrak{a}}+1},\dots,v_{\mathsf{L}_{\mathfrak{a}}+\mathsf{L}_{\mathfrak{b}}})^{\intercal}
\;,
\qquad
\mathfrak{c}(v)\;=\; 
(v_{\mathsf{L}_{\mathfrak{a}}+\mathsf{L}_{\mathfrak{b}}+1},\dots,v_{\mathsf{L}+1})^{\intercal}
\;.
$$
Moreover, let us introduce the \textit{macroscopic gap} $\gamma=\gamma\left(\mathcal{R},\mathsf{L}_{\mathfrak{b}},\mathsf{L}_{\mathfrak{c}}\right)$ between the upper and lower parts by
\begin{equation}
\label{eq-GammaDef}
\gamma
\;=\;
\min\left\{1\,,\,\frac{\kappa^2_{\mathsf{L}_{\mathfrak{c}}}}{\kappa^2_{\mathsf{L}_{\mathfrak{b}}+\mathsf{L}_{\mathfrak{c}}+1}}\,-\,1\right\}
\;\in\;\left[0,1\right]
\;.
\end{equation}
Note that the macroscopic gap $\gamma$ is positive provided that $\kappa_{\mathsf{L}_{\mathfrak{c}}}>\kappa_{\mathsf{L}_{\mathfrak{b}}+\mathsf{L}_{\mathfrak{c}}+1}$. Now the deviation of the random path $(v_n)_{n\in\mathbb{N}}$ defined by~\eqref{dynamics_sphere} and~\eqref{random_matrix} from the attractive part \mbox{$\{0\}^{\mathsf{L}_{\mathfrak{a}}}\times\mathbb{S}^{\mathsf{L}_{\mathfrak{b}}+\mathsf{L}_{\mathfrak{c}}-1}$} of the phase space can be measured as the norm of the upper part $\|\mathfrak{a}(v_N)\|$. The main result provides a quantitative bound on the expectation value of $\|\mathfrak{a}(v_N)\|^2$ for sufficiently large $N$ when the expectation is taken over the randomness contained in $\Tt_n$ for $n=1,\ldots,N$.

\begin{theo}\label{main_result}
Suppose that the i.i.d. $r_n\not\equiv 0$ are $[0,1]$-valued and that the i.i.d. $U_n$ are Haar distributed on $\textnormal{O}(\mathsf{L}+1)$. Furthermore suppose $\left(\mathsf{L}_{\mathfrak{a}},\mathsf{L}_{\mathfrak{b}}\right)\neq (1,1)$ and $\gamma>0$. Then, for all $0<\lambda\leq \frac{1}{4}$ there exist \mbox{$N_0= N_0(\mathsf{L},\mathsf{L}_{\mathfrak{c}},\lambda)\in\mathbb{N}$} such that 
\begin{equation}
\label{main_result_inequality}
\mathbb{E}\left\|\mathfrak{a}(v_N)\right\|^2
\;\leq \;2
\left(\frac{\mathsf{L}+1}{\mathsf{L}_{\mathfrak{a}}+\mathsf{L}_{\mathfrak{b}}}
\right)^{\frac{\mathsf{L}_{\mathfrak{a}}+\mathsf{L}_{\mathfrak{b}}-2}{\mathsf{L}_{\mathfrak{c}}+2}}
\left(\frac{6}{\gamma}\,\frac{\mathsf{L}_{\mathfrak{a}}}{\mathsf{L}_{\mathfrak{c}}}\,\lambda^2\right)^{\frac{\mathsf{L}_{\mathfrak{c}}}{2+\mathsf{L}_{\mathfrak{c}}}}
\end{equation}
for all $N\geq N_0$ and $v_0\in\mathbb{S}^{\mathsf{L}}$.
\end{theo}

Using the invariance property of the Furstenberg measure $\mu_{r,\lambda}$, one deduces the following

\begin{coro}
\label{coro-main_result}
Under the same hypothesis as in Theorem~\ref{main_result}, 
\begin{equation}
\label{eq-Coro-inequality}
\int \textnormal{d}\mu_{r,\lambda}(v)\,\left\|\mathfrak{a}(v)\right\|^2
\;\leq \;2
\left(\frac{\mathsf{L}+1}{\mathsf{L}_{\mathfrak{a}}+\mathsf{L}_{\mathfrak{b}}}
\right)^{\frac{\mathsf{L}_{\mathfrak{a}}+\mathsf{L}_{\mathfrak{b}}-2}{\mathsf{L}_{\mathfrak{c}}+2}}
\left(\frac{6}{\gamma}\,\frac{\mathsf{L}_{\mathfrak{a}}}{\mathsf{L}_{\mathfrak{c}}}\,\lambda^2\right)^{\frac{\mathsf{L}_{\mathfrak{c}}}{2+\mathsf{L}_{\mathfrak{c}}}}
\;.
\end{equation}
\end{coro}

The estimates \eqref{main_result_inequality} and  \eqref{eq-Coro-inequality} strongly differ from the behavior for $\Rr=\one$ where no hyperbolicity is present. Then $\mathbb{E}\|\mathfrak{a}(v_N)\|^2\sim \mathsf{L}_{\mathfrak{a}}\mathsf{L}^{-1}$ holds for large $N$ independent of $\lambda>0$ which just reflects the equidistribution of the random dynamics on all channels (this follows from Proposition~\ref{prop-RPP} below). To us, the most interesting regime is that of large $\mathsf{L}_{\mathfrak{a}}$, $\mathsf{L}_{\mathfrak{b}}$ and $\mathsf{L}_{\mathfrak{c}}$, say all a fraction of $\mathsf{L}$, and  of $\gamma$ of the order of  $1$ (but possibly less than $1$). Then the r.h.s. in \eqref{main_result_inequality} and  \eqref{eq-Coro-inequality} is approximately proportional to $\lambda^2$ which is the expected behavior. Indeed, the random kicks of order $\lambda$ are uniform and thus do not distinguish between channels, and hence the drift into each channel is given by their variance or $\lambda^2$, so that $\mathbb{E}\|\mathfrak{a}(v)\|^2$ should be of the order $\lambda^2$ times the proportion $\mathsf{L}_{\mathfrak{a}}\mathsf{L}^{-1}$ of channels in $\mathfrak{a}(v)$.

\vspace{.2cm}

Our main motivation for the present study are potential applications to the field of discrete random Schr\"odinger operators like the Anderson model, see \cite{BL,CL,AW} for general mathematical background information. Little is known rigorously about the so-called \textit{weak localization regime} of such operators in space dimension higher than or equal to $3$. In this regime, the eigenfunctions are not expected to be exponentially localized and the quantum dynamics is believed to be diffusive like in a Brownian motion.  Furthermore, random matrix theory is expected to provide a good description of the eigenvalues and eigenfunctions locally in space and within a suitable range of energies. In infinite volume the spectral measures likely have an absolutely continuous component.  The approach to this problem closest to the present study is the transfer matrix method. It allows to construct (generalized) eigenfunctions and Green functions of finite volume approximations. Best understood is then the quasi-one-dimensional limit in which one has strong Anderson localization, that is, pure-point spectrum with exponentially localized eigenfunctions with a rate called the \textit{inverse localization length} \cite{BL,CL,GM}. In a perturbative regime of small coupling of the randomness, one can calculate this localization length \cite{SB1,RS} and, more generally, the whole Lyapunov spectrum \cite{RS1,SS} provided the random dynamics of the transfer matrices is well understood. For such systems, one can also derive flow equations for the finite volume growth exponents, the so-called DMPK-equations \cite{BaR,VV,SV}. Beneath these works, only \cite{SB1,SV} address the hyperbolic character of the unperturbed dynamics (corresponding to the $\Rr$ above), however, only in the regime $\lambda\ll\delta\mathcal{R}_i$ of very small randomness \cite{SB1} or even a randomness vanishing at a suitable rate in the system size, namely the number of random matrices $\Tt_n$ involved \cite{SV}. 

\vspace{.15cm}

In order to apply the results of this paper (notably Theorem~\ref{main_result}) to the transfer matrices of the Anderson model and extract relevant information on its eigenfunctions, several non-trivial extensions have to be worked out. First of all, the transfer matrices at real energies have a symplectic symmetry that has to be implemented and then leads, in particular, to a supplementary symmetry in the Lyapunov spectrum. This can be done as in \cite{BL,SB1,LSS}. Then one has to consider the dynamics not only on unit vectors, but rather on the whole flag manifold \cite{BL,SB1}. Furthermore, while the transfer matrices can be brought in the form \eqref{random_matrix} \cite{SB1}, the random matrices $U_n$ stemming from the Anderson model are not Haar distributed and contain much fewer random entries. In the quasi-one-dimensional regime, this can be dealt with using commutator methods, see \cite{GM} and \cite{SS} for a perturbative result when $\Rr$ is elliptic, that is, of unit norm.

\vspace{.1cm}

Theorem~\ref{main_result} also has some short-comings by itself. First of all, it and its proof do not provide a good quantitative estimate on $N_0$. Furthermore, the proof does not readily transpose to the case where $\mathbf{1}+\lambda rU$ is replaced by $\exp(\lambda rU)$. Actually, many of the arguments below depend heavily on geometric considerations and explicit calculations exploring formulas for averages over the Haar measure. 

\vspace{.3cm}

\noindent {\bf Acknowledgements} We thank Andreas Knauf for many discussions as well as helpful and constructive comments. F.~D. received funding from the \textit{Studienstiftung des deutschen Volkes}. This work was also supported by the DFG.

\section{Outline of the proofs of Theorems~\ref{ac_distribution} and \ref{main_result}}

Throughout the remainder of the paper we assume that $\lambda\in(0,1)$ and that $(r_n)_{n\in\mathbb{N}}$ are i.i.d., $[0,1]$-valued and satisfy $\mathbb{P}(r_n=0)<1$. Furthermore $(U_n)_{n\in\mathbb{N}}$ are supposed to be $\textnormal{O}(\mathsf{L}+1)$-valued and i.i.d. according to the Haar measure. The one-dimensional Lebesgue measure will be denoted by $\textbf{x}$. We also abbreviate absolutely continuous and absolute continuity by a.c..

\begin{lemma}
\label{lem-0}
The random variable \mbox{\small$\left\langle v,(\mathbf{1}+\lambda rU)\cdot v\right\rangle$} is \mbox{\footnotesize $\big[\sqrt{1-\lambda^2},1\big]$}-valued for all $v\in\mathbb{S}^{\mathsf{L}}$. Moreover, its distribution is independent of $v\in\mathbb{S}^{\mathsf{L}}$.
\end{lemma}

Let us denote the Borel probability distribution of \mbox{\small$\left\langle v,(\mathbf{1}+\lambda rU)\cdot v\right\rangle$} by 
$$
\varpi_{r,\lambda}(A)
\;=\;
\mathbb{P}(\mbox{\small$\left\langle v,(\mathbf{1}+\lambda rU)\cdot v\right\rangle$}\in A)
\;,
\qquad
A\in\mathscr{B}(\mbox{\footnotesize $\big[\sqrt{1-\lambda^2},1\big]$})
\;.
$$
The aim of the next lemma is to analyze its canonical decomposition into pure-point, singular continuous and absolutely continuous component:
$$
\varpi_{r,\lambda}
\;=\;
\varpi_{r,\lambda}^{\textnormal{pp}}\;+\;\varpi_{r,\lambda}^{\textnormal{sc}}\;+\;\varpi_{r,\lambda}^{\textnormal{ac}}
\;.
$$

\begin{lemma}
\label{lem-1}
One has  $\varpi_{r,\lambda}^{\textnormal{pp}}=\mathbb{P}(r=0)\textnormal{ }\delta_1$ and $\varpi_{r,\lambda}^{\textnormal{sc}}=0$.
\end{lemma}

The next lemma states an elementary invariance property.

\begin{lemma}
\label{lem-2}
For all $v\in\mathbb{S}^{\mathsf{L}}$ the random variable $(\mathbf{1}+\lambda rU)\cdot v$ is distributed axially symmetrically w.r.t. $v$. More precisely, for all Borel subsets $A\in\mathscr{B}(\mathbb{S}^{\mathsf{L}})$, any orthogonal $V\in \textnormal{O}(\mathsf{L}+1)$ and all pairs $(\mathscr{V},v)\in\textnormal{O}(\mathsf{L}+1)\times\mathbb{S}^{\mathsf{L}}$ with $\mathscr{V}v=v$, one has
\begin{align}
\label{ac_distribution_8first}
\mathbb{P}\left((\mathbf{1}+\lambda rU)\cdot v\in A\right)
&
\;=\;
\mathbb{P}\left((\mathbf{1}+\lambda rU)\cdot V v\in VA\right)
\\
&
\label{ac_distribution_8}
\;=\;
\mathbb{P}\left((\mathbf{1}+\lambda rU)\cdot v\in \mathscr{V}A\right)\,.
\end{align}
\end{lemma}

Lemmata~\ref{lem-1} and \ref{lem-2} allow to consider the Borel probability distribution on $\mathbb{S}^{\mathsf{L}}$ of the random variable $(\mathbf{1}+\lambda rU)\cdot v$:
$$
\varrho_{r,\lambda,v}(A)
\;=\;
\mathbb{P}((\mathbf{1}+\lambda rU)\cdot v\in A)\;,
\qquad
A\in\mathscr{B}(\mathbb{S}^{\mathsf{L}})
\;.
$$
The next lemma analyzes its canonical decomposition $\varrho_{r,\lambda,v}=\varrho_{r,\lambda,v}^{\textnormal{pp}}+\varrho_{r,\lambda,v}^{\textnormal{sc}}+\varrho_{r,\lambda,v}^{\textnormal{ac}}$ w.r.t. $\nu_{\mathsf{L}}$.

\begin{lemma}
\label{lem-2a}
Let $v,w\in\mathbb{S}^{\mathsf{L}}$. Then
\begin{equation}
\label{lem-2a-ppsc}
\varrho_{r,\lambda,v}^{\textnormal{pp}}
\,=\,
\mathbb{P}(r=0)\textnormal{ }\delta_v
\;,
\qquad
\varrho_{r,\lambda,v}^{\textnormal{sc}}\,=\,0
\;,
\end{equation}
and the Radon-Nikodym derivative of the absolutely continuous part $\varrho_{r,\lambda,v}^{\textnormal{ac}}$ w.r.t. $\nu_{\mathsf{L}}$ obeys the following symmetry property:
\begin{equation}
\label{lem-2a-ac}
\frac{\textnormal{d}\varrho^{\textnormal{ac}}_{r,\lambda,v}}{\textnormal{d}\nu_{\mathsf{L}}}(w)
\;=\;
\frac{\textnormal{d}\varrho^{\textnormal{ac}}_{r,\lambda,w}}{\textnormal{d}\nu_{\mathsf{L}}}(v)\,.
\end{equation}
\end{lemma}


The final preparatory result involves the deterministic hyperbolic part $\mathcal{R}\cdot$ of the dynamics.

\begin{lemma}
\label{lem-4}
The absolute continuity of Borel measures on $\mathbb{S}^{\mathsf{L}}$ w.r.t. $\nu_{\mathsf{L}}$ is preserved under~$(\mathcal{R}\cdot)_*$.
\end{lemma}

Once all these lemmata are proved (once again, see Section~\ref{sec-ProofDetails}), it is possible to complete the proof of the first part of Theorem~\ref{ac_distribution}, namely to prove the absolute continuity stated therein.

\vspace{.2cm}

\noindent{\bf Proof} of  Theorem~\ref{ac_distribution}.
For $n\in\mathbb{N}$, let us denote the distribution of $v_n$ for some given initial condition $v_0\in\mathbb{S}^{\mathsf{L}}$ by $\varsigma_n$. It can be computed iteratively by
\begin{equation}
\label{proof-ac-distribution-eq-1}
\varsigma_n
\;=\;
\int_{\mathbb{S}^{\mathsf{L}}}\textnormal{d}\varsigma_{n-1}(w)\; ((\mathcal{R}\cdot)_*(\varrho_{r,\lambda,w}))(\cdot)\;,
\qquad
\varsigma_0\;=\;\delta_{v_0}\;.
\end{equation}
Now, let $\aleph\in\mathscr{B}(\mathbb{S}^{\mathsf{L}})$ be a $\nu_{\mathsf{L}}$-nullset. Then, $\aleph$ is also an $((\mathcal{R}\cdot)_*(\varrho_{r,\lambda,w}^{\textnormal{ac}}))$-nullset by Lemma~\ref{lem-4}. Therefore \eqref{proof-ac-distribution-eq-1} combined with~\eqref{lem-2a-ppsc} implies that
\begin{align}
\label{proof-ac-distribution-eq-3}
\begin{split}
\varsigma_n(\aleph)
&
\;=\;
\int_{\mathbb{S}^{\mathsf{L}}}\textnormal{d}\varsigma_{n-1}(w)\;((\mathcal{R}\cdot)_*(\varrho_{r,\lambda,w}))(\aleph)
\\
&
\;=\;
\int_{\mathbb{S}^{\mathsf{L}}}\textnormal{d}\varsigma_{n-1}(w)\;((\mathcal{R}\cdot)_*(\varrho_{r,\lambda,w}^{\textnormal{pp}}))(\aleph)
\\
&
\;=\;
\mathbb{P}(r=0)
\;
\int_{\mathbb{S}^{\mathsf{L}}}\textnormal{d}\varsigma_{n-1}(w)\;((\mathcal{R}\cdot)_*(\delta_{w}))(\aleph)
\\
&
\;=\;
\mathbb{P}(r=0)
\;
((\mathcal{R}\cdot)_*(\varsigma_{n-1}))(\aleph)
\;.
\end{split}
\end{align}
By iteratively applying~\eqref{proof-ac-distribution-eq-3} from $n=1$ to some $N\geq 1$ and inserting $\varsigma_0=\delta_{v_0}$ one obtains
\begin{equation}
\label{proof-ac-distribution-eq-4}
\varsigma_N(\aleph)
\;=\;
\mathbb{P}(r=0)^N
\;
((\mathcal{R}^N\cdot)_*(\varsigma_{0}))(\aleph)
\;=\;
\mathbb{P}(r=0)^N
\;
((\mathcal{R}^N\cdot)_*(\delta_{v_0}))(\aleph)
\,.
\end{equation}
An iterative application of the invariance property of the Furstenberg measure $\mu_{r,\lambda}$ yields
\begin{align*}
\mu_{r,\lambda}
\;=\;
\int\limits_{\mathbb{S}^{\mathsf{L}}}\textnormal{d}\mu_{r,\lambda}(v)\;\varsigma_N\big|_{v_0=v}\,,
\end{align*}
which implies together with~\eqref{proof-ac-distribution-eq-4} that
\begin{align}
\label{proof-ac-distribution-eq-5}
\mu_{r,\lambda}(\aleph)
\;=\;
\int\limits_{\mathbb{S}^{\mathsf{L}}}\textnormal{d}\mu_{r,\lambda}(v)\;\varsigma_N(\aleph)\big|_{v_0=v}
\;=\;
\mathbb{P}(r=0)^N
\;
((\mathcal{R}^N\cdot)_*(\mu_{r,\lambda}))(\aleph)
\;\leq\;
\mathbb{P}(r=0)^N
\,.
\end{align}
Due to the assumption $r\not\equiv 0$, \textit{i.e.}, $\mathbb{P}(r=0)<1$, the absolute continuity of $\mu_{r,\lambda}$ w.r.t. $\nu_{\mathsf{L}}$ follows from~\eqref{proof-ac-distribution-eq-5} in the limit $N\rightarrow\infty$. If $\mathbb{P}(r=0)=0$ holds, then~\eqref{proof-ac-distribution-eq-4} implies that even the distribution of $v_N$ is absolutely continuous w.r.t. $\nu_{\mathsf{L}}$ for all $N\geq 1$.

\vspace{.1cm}

The penultimate statement of Theorem~\ref{ac_distribution} is Lemma~\ref{large-microscopic-gap} below. Let us now focus on the last claim, namely the fact that the support of the Furstenberg measure is the whole sphere $\SM^{\mathsf{L}}$ if $\lambda> \delta\mathcal{R}_i$ for all $i$. More precisely, we show that $\mu_{r,\lambda}(B_{\epsilon}(w))>0$ holds for every ball of radius $\epsilon>0$ around any arbitrary point $w\in\mathbb{S}^{\mathsf{L}}$. For this purpose, let us pick some $u\in\textnormal{supp}(\mu_{r,\lambda})$. In view of Lemma~\ref{reach_point} (see also below), there exists a path of finite length from $u$ to $w$, {\it i.e.}, there exists $N\in\mathbb{N}$ and $\left\{s_n\right\}_{n=1}^N\subset \textnormal{supp}(r)$ and $\left\{\mathscr{U}_n\right\}_{n=1}^N\subset\textnormal{O}(\mathsf{L}+1)=\textnormal{supp}(U)$ such that
$$
w\;=\;\prod\limits_{n=1}^N\mathcal{R}(\mathbf{1}+\lambda s_n\mathscr{U}_n)\cdot u
$$
holds. Obviously, the event
$$
\Big\|\prod\limits_{n=1}^N\mathcal{R}(\mathbf{1}+\lambda r_nU_n)-\prod\limits_{n=1}^N\mathcal{R}(\mathbf{1}+\lambda s_n\mathscr{U}_n)\Big\|
\;<\;\zeta
$$
has positive probability for all $\zeta>0$ and the map $(A,v)\mapsto A\cdot v$ is continuous. Therefore, there exists some $\xi>0$ such that $\mathbb{P}(v_N\in B_{\epsilon}(w))>0$ for all $v_0\in B_{\xi}(u)$. Now every ball $B_{\xi}(u)$ of radius $\xi>0$ around $u\in\textnormal{supp}(\mu_{r,\lambda})$ satisfies $\mu_{r,\lambda}(B_{\xi}(u))>0$. This allows to infer that
$$
\mu_{r,\lambda}(B_{\epsilon}(w))
\;=\;
\int\limits_{\mathbb{S}^{\mathsf{L}}}\textnormal{d}\mu_{r,\lambda}(v_0)\textnormal{ }\mathbb{P}(v_N\in B_{\epsilon}(w))
\;\geq\; 
\int\limits_{B_{\xi}(u)}\textnormal{d}\mu_{r,\lambda}(v_0)\textnormal{ }\mathbb{P}(v_N\in B_{\epsilon}(w))
\;>\;
0\;,
$$
which proves the claim.
\hfill $\Box$

\begin{lemma}\label{large-microscopic-gap}
If $\lambda< 2^{-4}\min\{\delta\mathcal{R}_i,\mbox{\small $\frac{1}{2}$}\}$ for some $i=1,\ldots,\mathsf{L}$, then $\textnormal{supp}(\mu_{r,\lambda})\neq\mathbb{S}^{\mathsf{L}}$.
\end{lemma}

\begin{lemma}
\label{reach_point}
Suppose that $\lambda > \max_{i=1,\ldots,\mathsf{L}}\delta\mathcal{R}_i$ and that $1\in \textnormal{supp}(r)$. Then for every couple $u,w\in\mathbb{S}^{\mathsf{L}}$ there exist $N\in\mathbb{N}$ and $s_1,\ldots,s_N\in \textnormal{supp}(r)$ and $\mathscr{U}_1,\ldots,\mathscr{U}_N\in\textnormal{O}(\mathsf{L}+1)$ such that
$$
w\;=\;
\prod\limits_{n=1}^N\mathcal{R}(\mathbf{1}+\lambda s_n\mathscr{U}_n)\cdot u\;.
$$
\end{lemma}

Next let us outline the proof of Theorem~\ref{main_result}. It will be useful to split each $\mathcal{T}_n$ into the unperturbed, deterministic action $\mathcal{R}$ and a random perturbation $\mathbf{1}+\lambda rU$, and analyze the action of both factors separately. The unperturbed action $\mathcal{R}\cdot$ leads to a decrease of the norm of the upper part and an increase of the norm of the lower part. More precisely, provided that $\kappa_{\mathsf{L}_{\mathfrak{c}}}>\kappa_{\mathsf{L}_{\mathfrak{b}}+\mathsf{L}_{\mathfrak{c}}+1}$, one has for any $v\in\mathbb{S}^{\mathsf{L}}$ obeying $\left\|\mathfrak{a}(v)\right\|\neq 0\neq \left\|\mathfrak{c}(v)\right\|$ the bounds
\begin{align}
\label{decrease_a_increase_c}
\left\|\mathfrak{a}(\mathcal{R}\cdot v)\right\|
\;<\;
\left\|\mathfrak{a}(v)\right\|\;,
\qquad\left\|\mathfrak{c}(\mathcal{R}\cdot v)\right\|
\;>\;
\left\|\mathfrak{c}(v)\right\|
\;.
\end{align}
The former inequality is now strengthened. 

\begin{lemma}\label{deterministic_estimate}
For all $v\in\mathbb{S}^{\mathsf{L}}$,
\begin{align}\label{deterministic_estimate_inequality}
\left\|\mathfrak{a}\left(\mathcal{R}\cdot v\right)\right\|^2
\;\leq\; 
\left(1-\left\|\mathfrak{c}(v)\right\|^2\frac{\gamma}{2}\right)\left\|\mathfrak{a}(v)\right\|^2
\;.
\end{align}
\end{lemma}

This implies that the unperturbed dynamics obeys
\begin{align*}
\lim_{N\rightarrow\infty}\,\mathfrak{a}(\mathcal{R}^N\cdot v_0)
\;=\;
0
\end{align*}
if $\left\|\mathfrak{c}(v_0)\right\|>0$ and $\kappa_{\mathsf{L}_{\mathfrak{c}}}>\kappa_{\mathsf{L}_{\mathfrak{b}}+\mathsf{L}_{\mathfrak{c}}+1}$. The random perturbation, on the other hand, may augment $\|\mathfrak{a}(v)\|$. However, in expectation this growth is bounded by a term of order $\mathcal{O}(\lambda^2)$.

\begin{lemma}\label{bound_average_drift}
Let $\lambda\in (0,\frac{1}{4}]$ and $\mathsf{L}\geq 3$. Then for all $v\in\mathbb{S}^{\mathsf{L}}$,
\begin{align}
\label{bound_average_drift_inequality}
\mathbb{E}\left\|\mathfrak{a}\left(\left(\mathbf{1}+\lambda rU\right)\cdot v\right)\right\|^2
\;\leq\;
\left\|\mathfrak{a}(v)\right\|^2\,+\,\lambda^2 \frac{3\,\mathsf{L}_{\mathfrak{a}}}{\mathsf{L}+1}
\;.
\end{align}
\end{lemma}

At first glance, it may now appear straightforward to prove upper bounds on $\mathbb{E} \|\mathfrak{a}(v_n)\|^2$ for large $N$ by combining Lemmata~\ref{deterministic_estimate} and \ref{bound_average_drift}. An iterative application turns out to be more involved, however. The core task is to deal with the expectation value of products $\left\|\mathfrak{a}(v_n)\right\|^2\left\|\mathfrak{c}(v_n)\right\|^2$ in~\eqref{deterministic_estimate_inequality}. This is tackled by the following elementary lemma.

\begin{lemma}\label{get_rid_of_unlikely_cases}
$\mathbb{S}^{\mathsf{L}}$-valued random variables $u$ with arbitrary distribution satisfy for all $\delta\in [0,1]$
\begin{align}\label{get_rid_of_unlikely_cases_inequality}
\mathbb{E}\,\left\|\mathfrak{a}(u)\right\|^2\left\|\mathfrak{c}(u)\right\|^2
\;\geq\; 
\delta\left[\mathbb{E}\,\left\|\mathfrak{a}(u)\right\|^2\,-\,\mathbb{P}\mbox{\small $\left(\left\|\mathfrak{c}(u)\right\|^2<\delta\right)$}\right]
\;.
\end{align}
\end{lemma}

Consequently the next aim is to bound
\begin{align}
\label{bound_by_random_phase_system_term_to_be_bounded}
\mathbb{P}\big(\left\|\mathfrak{c}\left((\mathbf{1}+\lambda rU)\cdot  v_n\right)\right\|^2\leq \delta\big)
\end{align}
from above so that inequalities \eqref{deterministic_estimate_inequality} and \eqref{get_rid_of_unlikely_cases_inequality} can be used.  This turns out to be possible by comparing the random dynamics \eqref{dynamics_sphere} generated by \eqref{random_matrix} with the random dynamics generated by $\one+\lambda r_n U_n$ instead of $\Tt_n$, that is, the case of $\mathcal{R}=\mathbf{1}$ which has no hyperbolicity. The comparison of the cumulative distribution function \eqref{bound_by_random_phase_system_term_to_be_bounded} under these two random dynamics is based on the next result.

\begin{proposi}
\label{prop-StochOd}
Let $\left(\mathsf{L}_{\mathfrak{a}},\mathsf{L}_{\mathfrak{b}}\right)\neq (1,1)$ and $v,w\in\mathbb{S}^{\mathsf{L}}$ be such that $\left\|\mathfrak{c}(v)\right\|\geq \left\|\mathfrak{c}(w)\right\|$. For all $\epsilon\in [0,1]$ and $\lambda\in (0,\frac{1}{4}]$, one then has
\begin{align}
\label{basic_stochastic_order_inequality}
\mathbb{P}\big(\left\|\mathfrak{c}\left((\mathsf{1}+\lambda rU)\cdot v\right)\right\|\leq 
\epsilon\big)
\;\leq\;
\mathbb{P}\big(\left\|\mathfrak{c}\left((\mathsf{1}+\lambda rU)\cdot w\right)\right\|\leq\epsilon\big)
\;.
\end{align}
\end{proposi}

\vspace{.2cm}

\noindent {\bf Remark}
Since $\left\|\mathfrak{c}\left((\mathsf{1}+\lambda rU)\cdot v\right)\right\|$ and $\left\|\mathfrak{c}\left((\mathsf{1}+\lambda rU)\cdot w\right)\right\|$ are $\mathbb{R}$-valued, the validity of~\eqref{basic_stochastic_order_inequality} for all $\epsilon\in [0,1]$ is equivalent to the stochastic order
\begin{align}
\mathbb{P}\big(\left\|\mathfrak{c}\left((\mathsf{1}+\lambda rU)\cdot v\right)\right\| \in \cdot\big)
\;\geq_{\textnormal{st}}\;
\mathbb{P}\big(\left\|\mathfrak{c}\left((\mathsf{1}+\lambda rU)\cdot w\right)\right\| \in \cdot\big)
\;,
\end{align}
as defined, {\it e.g.}, in Section 17.7 of \cite{Kle}.
\hfill $\diamond$

\vspace{.2cm}

Now one can iteratively combine the second part of \eqref{decrease_a_increase_c} and Proposition~\ref{prop-StochOd}. For ordered products, we use the following notation:
$$
\prod_{i=j}^kF_i
\;=\;
\begin{cases}F_k\cdots F_j\;,&\;\; j\leq k\;,\\ \mathbf{1}\;, & \;\;j>k\;.\end{cases}
$$

\begin{coro}\label{bound_by_random_phase_system}
Let $\left(\mathsf{L}_{\mathfrak{a}},\mathsf{L}_{\mathfrak{b}}\right)\neq (1,1)$. Then for all $v\in\mathbb{S}^{\mathsf{L}}$, $\epsilon\in [0,1]$, $N\in\mathbb{N}$ and $\lambda\in (0,\frac{1}{4}]$, 
\begin{equation}
\label{bound_by_random_phase_system_inequality}
\mathbb{P}\Big(\Big\|\mathfrak{c}\big((\mathbf{1}+\lambda r_N U_N)\prod_{n=1}^{N-1}\mathcal{R}(\mathbf{1}+\lambda r_nU_n)\cdot v\big)\Big\|\leq \epsilon \Big)
\,\leq\,
\mathbb{P}\Big(\Big\|\mathfrak{c}\big(\prod_{n=1}^{N}(\mathbf{1}+\lambda r_nU_n)\cdot v\big)\Big\|\leq \epsilon \Big)
\;.
\end{equation}
\end{coro}

Corollary~\ref{bound_by_random_phase_system} allows to bound \eqref{bound_by_random_phase_system_term_to_be_bounded} by the r.h.s. of \eqref{bound_by_random_phase_system_inequality} with $\delta=\epsilon^2$. This r.h.s. can readily be estimated if one knows the 
invariant probability measure on $\mathbb{S}^{\mathsf{L}}$ under the dynamics $(\mathbf{1}+\lambda rU)\cdot$ (it is again unique and given by the Furstenberg measure). The following proposition shows that this invariant measure is equal to the normalized invariant surface measure $\nu_{\mathsf{L}}$ on $\mathbb{S}^{\mathsf{L}}$. In the terminology of \cite{SS,RS1} this means that the dynamics $(\mathbf{1}+\lambda rU)\cdot$ has the so-called \textit{random phase property}. 

\begin{proposi}
\label{prop-RPP}
For all $\lambda\in (0,\frac{1}{4}]$ and $h\in\mbox{\large $\mathtt{L}$}^{\infty}(\mathbb{S}^{\mathsf{L}})$, one has
\begin{align}
\label{RPP_equation}
\int_{\mathbb{S}^{\mathsf{L}}}\textnormal{d}\nu_{\mathsf{L}}(v)\;\mathbb{E}\,h\left((\mathbf{1}+\lambda rU)\cdot v\right)
\;=\;
\int_{\mathbb{S}^{\mathsf{L}}}\textnormal{d}\nu_{\mathsf{L}}(v)\;h(v)
\end{align}
\end{proposi}


At large $N$, the r.h.s. of \eqref{bound_by_random_phase_system_inequality} therefore approaches $\nu_{\mathsf{L}}(\{v\in\mathbb{S}^{\mathsf{L}}: \|\mathfrak{c}(v)\|^2<\delta\})$ (see~\cite{BL}, Part A, Theorem 4.3). Therefore the following geometric identity will be needed.

\begin{lemma}\label{probability_RPP}
For all $\delta\in [0,1]$,
\begin{equation}
\label{eq-SphereEst}
\nu_{\mathsf{L}}\left(\left\{v\in\mathbb{S}^{\mathsf{L}}: \left\|\mathfrak{c}(v)\right\|^2
<\delta\right\}\right)
\;=\;
\frac{\Gamma(\tfrac{\mathsf{L}+1}{2})}{
\Gamma(\tfrac{\mathsf{L}_{\mathfrak{c}}}{2})\Gamma(\tfrac{\mathsf{L}_{\mathfrak{a}}+\mathsf{L}_{\mathfrak{b}}}{2})}
\;
\int^{\delta}_0
\textnormal{d}\textnormal{\textbf{x}}(x)
\;
x^{\frac{\mathsf{L}_{\mathfrak{c}}}{2}-1}(1-x)^{\frac{\mathsf{L}_{\mathfrak{a}}+\mathsf{L}_{\mathfrak{b}}}{2}-1}
\;,
\end{equation}
which just means that $\|\mathfrak{c}(v)\|^2$ is distributed according to the beta distribution with parameters $(\frac{\mathsf{L}_{\mathfrak{a}}+\mathsf{L}_{\mathfrak{b}}}{2},\frac{\mathsf{L}_{\mathfrak{c}}}{2})$.
For $\left(\mathsf{L}_{\mathfrak{a}},\mathsf{L}_{\mathfrak{b}}\right)\neq (1,1)$ this can, moreover, be bounded as follows:
\begin{align}
\label{probability_RPP_inequality}
\nu_{\mathsf{L}}\left(\left\{v\in\mathbb{S}^{\mathsf{L}}: \left\|\mathfrak{c}(v)\right\|^2
<\delta\right\}\right)
\;\leq\;
\Big(\frac{\mathsf{L}+1}{\mathsf{L}_{\mathfrak{a}}+\mathsf{L}_{\mathfrak{b}}}\Big)^{\frac{\mathsf{L}_{\mathfrak{a}}+\mathsf{L}_{\mathfrak{b}}}{2}-1}
\Big(\frac{\mathsf{L}+1}{\mathsf{L}_{\mathfrak{c}}}\,\delta\Big)^{\frac{\mathsf{L}_{\mathfrak{c}}}{2}}
\Big(1-\frac{\delta}{6}\Big)
\;.
\end{align} 
\end{lemma}

The following Corollary~\ref{exploit_RPP} combines Proposition~\ref{prop-RPP} and Lemma~\ref{probability_RPP} and concludes the transient focus on the special case of $\mathcal{R}=\mathbf{1}$.

\begin{coro}\label{exploit_RPP}
Let $\left(\mathsf{L}_{\mathfrak{a}},\mathsf{L}_{\mathfrak{b}}\right)\neq (1,1)$ and $\delta\in (0,1)$.  Then there exist $\tilde{N_0}= \tilde{N_0}(\mathsf{L},\mathsf{L}_{\mathfrak{c}},\delta)\in\mathbb{N}$ and $\eta=\eta(\mathsf{L},\mathsf{L}_{\mathfrak{c}},\delta)>0$ such that
\begin{align}
\label{exploit_RPP_inequality}
\mathbb{P}\Big(\Big\| \mathfrak{c}\Big(\prod_{n=1}^{N}(\mathbf{1}+\lambda r_nU_n)\cdot v\Big)\Big\|^2<\delta
\Big)
\;\leq\; 
\Big(\frac{\mathsf{L}+1}{\mathsf{L}_{\mathfrak{a}}+\mathsf{L}_{\mathfrak{b}}}\Big)^{\frac{\mathsf{L}_{\mathfrak{a}}+\mathsf{L}_{\mathfrak{b}}}{2}-1}
\Big(\frac{\mathsf{L}+1}{\mathsf{L}_{\mathfrak{c}}}\,\delta\Big)^{\frac{\mathsf{L}_{\mathfrak{c}}}{2}}
\,-\,\eta
\end{align}
holds for all $N\geq \tilde{N_0}$ and $v\in\mathbb{S}^{\mathsf{L}}$.
\end{coro}

Lemmata~\ref{deterministic_estimate},~\ref{bound_average_drift},~\ref{get_rid_of_unlikely_cases} and Corollaries~\ref{bound_by_random_phase_system} and~\ref{exploit_RPP} now allow to conclude.

\vspace{.2cm}

\noindent {\bf Proof} of Theorem~\ref{main_result}. 
Let $\delta\in (0,1)$ and $\eta=\eta(\mathsf{L},\mathsf{L}_{\mathfrak{c}},\delta)>0$ and $\tilde{N_0}=\tilde{N_0}(\mathsf{L},\mathsf{L}_{\mathfrak{c}},\delta)\in\mathbb{N}$ be as in Corollary~\ref{exploit_RPP}. Moreover, let us choose $\tilde{N}\geq \tilde{N_0}$. Then Lemmata~\ref{deterministic_estimate},~\ref{bound_average_drift},~\ref{get_rid_of_unlikely_cases} and Corollaries~\ref{bound_by_random_phase_system} and~\ref{exploit_RPP} imply the estimate
\begin{align*}
& \mathbb{E}\,\|\mathfrak{a}(v_{\tilde{N}+1})\|^2
\;=\;
\mathbb{E}\,\|\mathfrak{a}(\mathcal{R}(\mathbf{1}+\lambda r_{\tilde{N}+1}U_{\tilde{N}+1})\cdot v_{\tilde{N}})\|^2
%
\\
&
\;\;\leq\; 
\big(1-\frac{\gamma\delta}{2}\big)\;\mathbb{E}\,
\|\mathfrak{a}((\mathbf{1}+\lambda r_{\tilde{N}+1}U_{\tilde{N}+1})\cdot v_{\tilde{N}})\|^2
 +\frac{\gamma\delta}{2}\;\mathbb{P}\big(\|\mathfrak{c}((\mathbf{1}+\lambda r_{\tilde{N}+1}U_{\tilde{N}+1})\cdot v_{\tilde{N}})\|^2<\delta\big)
\\
&
\;\;\leq\; 
\big(1-\frac{\gamma\delta}{2}\big)\Big[\mathbb{E}
\,\|\mathfrak{a}(v_{\tilde{N}})\|^2
+\lambda^2\,\frac{3\,\mathsf{L}_{\mathfrak{a}}}{\mathsf{L}+1}\Big]
\;+\;
\Big[
\Big(\frac{\mathsf{L}+1}{\mathsf{L}_{\mathfrak{a}}+\mathsf{L}_{\mathfrak{b}}}\Big)^{\frac{\mathsf{L}_{\mathfrak{a}}+\mathsf{L}_{\mathfrak{b}}}{2}-1}
\Big(\frac{\mathsf{L}+1}{\mathsf{L}_{\mathfrak{c}}}\,\delta\Big)^{\frac{\mathsf{L}_{\mathfrak{c}}}{2}}
\,-\,\eta
\Big]\frac{\gamma\delta}{2}
\\
&
\;\;\leq\;  
\big(1-\frac{\gamma\delta}{2}\big)\,\mathbb{E}
\,\|\mathfrak{a}(v_{\tilde{N}})\|^2
\;+\;
M_\delta
\;-\;\frac{\gamma\delta\eta}{2}\,,
\end{align*}
where 
$$
M_\delta
\;=\;
\lambda^2\,\frac{3\,\mathsf{L}_{\mathfrak{a}}}{\mathsf{L}+1}
\;+\;
\frac{\gamma\mathsf{L}_{\mathfrak{c}}}{2(\mathsf{L}+1)}
\Big(\frac{\mathsf{L}+1}{\mathsf{L}_{\mathfrak{a}}+\mathsf{L}_{\mathfrak{b}}}\Big)^{\frac{\mathsf{L}_{\mathfrak{a}}+\mathsf{L}_{\mathfrak{b}}}{2}-1}
\Big(\frac{\mathsf{L}+1}{\mathsf{L}_{\mathfrak{c}}}\,\delta\Big)^{\frac{\mathsf{L}_{\mathfrak{c}}}{2}+1}
\;.
$$
An iterative application of this inequality from $\tilde{N}=\tilde{N_0}$ to $N-1$ yields
\begin{align*}
\mathbb{E}\|\mathfrak{a}(v_{N})\|^2
& 
\;\leq\; 
\big(1-\
\frac{\gamma\delta}{2}\big)^{N-\tilde{N_0}}\mathbb{E}\|\mathfrak{a}(v_{\tilde{N_0}})\|^2
\;+\;
\big[M_\delta
\;-\;\frac{\gamma\delta\eta}{2}\big]
\sum_{\tilde{N}=\tilde{N_0}}^{N-1}
\big(1-\frac{\gamma\delta}{2}\big)^{\tilde{N}-\tilde{N_0}}
\\
&
\;\leq\;
\big(1-\frac{\gamma\delta}{2}\big)^{N-\tilde{N_0}}
\;+\;
\frac{2}{\gamma\delta}\;\big[M_\delta
\;-\;\frac{\gamma\delta\eta}{2}\big]
\end{align*}
for all $N\geq \tilde{N_0}$. Thus for all
$$
N\;\geq\; \tilde{N_0}\;+\;\frac{\log(\eta)}{\log\big(1-\frac{\gamma\delta}{2}\big)}
$$
one has
\begin{align}\label{upper-bound-delta}
\mathbb{E}\|\mathfrak{a}(v_{N})\|^2
\;\leq\;
\frac{2\,M_\delta}{\gamma\delta}
\;.
\end{align}
Now, the right side of~\eqref{main_result_inequality} is larger than $1$ if 
\begin{align*}
\mathtt{d}
\;=\;
\frac{\mathsf{L}_{\mathfrak{c}}}{\mathsf{L}+1}
\;
\Big(\frac{6\lambda^2\mathsf{L}_{\mathfrak{a}}}{\gamma \mathsf{L}_{\mathfrak{c}}}
\Big)^{\frac{2}{\mathsf{L}_{\mathfrak{c}}+2}}
\;
\Big(\frac{\mathsf{L}_{\mathfrak{a}}+\mathsf{L}_{\mathfrak{b}}}{\mathsf{L}+1}
\Big)^{\frac{\mathsf{L}_{\mathfrak{a}}+\mathsf{L}_{\mathfrak{b}}-2}{\mathsf{L}_{\mathfrak{c}}+2}}
\end{align*}
satisfies $\mathtt{d}\geq 1$. If this is violated, the choice $\delta=\mathtt{d}$ is possible and optimizes the order of the right side of~\eqref{upper-bound-delta} in $\lambda$ and proves \eqref{main_result_inequality}.
\hfill $\Box$

\section{Details of the proof of Theorem~\ref{ac_distribution}}
\label{sec-ProofDetails}

\noindent {\bf Proof} of Lemma~\ref{lem-0}.
The first item is obvious. As for the dependence of the distribution of \mbox{\small$\left\langle v,(\mathbf{1}+\lambda rU)\cdot v\right\rangle$} on $v\in\mathbb{S}^{L}$, let $w\in\mathbb{S}^{\mathsf{L}}$ and $\mathscr{W}\in\textnormal{O}(\mathsf{L}+1)$ be such that $\mathscr{W}w=v$. For all $s\in [0,1]$ and $\mathscr{U}\in \textnormal{O}(\mathsf{L}+1)$ one has
\begin{align*}
\left\langle v,(\mathbf{1}+\lambda s\mathscr{U})\cdot v\right\rangle
&
\;=\;  
\left\|(\mathbf{1}+\lambda s\mathscr{U})v\right\|^{-1}\; \left\langle v,(\mathbf{1}+\lambda s\mathscr{U})v\right\rangle 
\\
&
\;=\;
\left\|(\mathbf{1}+\lambda s\mathscr{W}^*\mathscr{U}\mathscr{W})w\right\|^{-1}\;
\left\langle w, (\mathbf{1}+\lambda s\mathscr{W}^*\mathscr{U}\mathscr{W}) w\right\rangle
\\
&
\;=\;
\left\langle w, (\mathbf{1}+\lambda s\mathscr{W}^*\mathscr{U}\mathscr{W})\cdot w\right\rangle
\;,
\end{align*}
but $\mathscr{W}^*U\mathscr{W}$ is distributed identically to $U$ due to the invariance of the Haar measure.
\hfill $\Box$

\vspace{.2cm}

\noindent {\bf Proof} of Lemma~\ref{lem-1}.
The normalized surface measure $\nu_{\mathsf{L}}$ on $\mathbb{S}^{\mathsf{L}}$ is equal to the push-forward $\left(\tau_v\right)_*(\theta_{\mathsf{L}})=\theta_{\mathsf{L}}\circ \left(\tau_v\right)^{-1}$ of the Haar measure $\theta_{\mathsf{L}}$ on $\textnormal{O}(\mathsf{L}+1)$ under the map $\tau_v:\textnormal{O}(\mathsf{L}+1)\rightarrow\mathbb{S}^{\mathsf{L}}$ given by $\tau_v( U)= Uv$, independently of the choice of $v\in\mathbb{S}^{\mathsf{L}}$ (see \cite{Mat}, Chapter 3). Considering, moreover, the projection  $\varsigma_v:\mathbb{S}^{\mathsf{L}}\rightarrow\mathbb{R}$ into the direction $v$ given by $\varsigma_v(w)=\langle v,w\rangle$, it is also known that the push-forward $\left(\varsigma_v\right)_*(\nu_{\mathsf{L}})=\nu_{\mathsf{L}}\circ \left(\varsigma_v\right)^{-1}$ of $\nu_{\mathsf{L}}$ is a.c. w.r.t. $\textbf{x}$ with a Radon-Nikodym density given by
\begin{align}
\label{ac_distribution_0}
\frac{\Gamma(\tfrac{\mathsf{L}+1}{2})}{\sqrt{\pi}\,\Gamma(\tfrac{\mathsf{L}}{2})}\;
\big(1-(\cdot)^2\big)^{\frac{\mathsf{L}}{2}-1}
\;\chi_{[-1,1]}
\;.
\end{align}
The action \eqref{eq-ActionDef} can be spelled out explicitly in terms of the random variable $Y=\left\langle v,Uv\right\rangle$ as
\begin{equation}
\label{eq-ActionFormula}
\left\langle v,(\mathbf{1}+\lambda rU)\cdot v\right\rangle 
\;=\;
\frac{1+\lambda rY}{\sqrt{1+2\lambda rY+\lambda^2r^2}}
\;.
\end{equation}
Thus let us denote the r.h.s. of~\eqref{eq-ActionFormula} by $G(r,Y)$. As $Y$ is distributed according to $\left(\varsigma_v\right)_*(\nu_{\mathsf{L}})$, it is a.c. w.r.t. $\textbf{x}$ on $[-1,1]$.  For $s\in (0,1]$, let  $H_{\pm}^s:\big[\sqrt{1-\lambda^2s^2},1\big]\to \mathbb{R}$ be the two inverse branches of $y\mapsto G(s,y)$. They are given by
$$
H_{\pm}^s(z)\;=\; \Big[z^2-1\pm z\big(z^2-1+\lambda^2s^2\big)^{\frac{1}{2}}\Big](\lambda s)^{-1}\;.
$$
For every $\textnormal{\textbf{x}}$-nullset $\aleph$ one thus has
$$
G(s,\cdot)^{-1}(\aleph)
\;=\;
\bigcup\limits_{\sigma=\pm}H_{\sigma}^s\big(\aleph\cap\big[\sqrt{1-\lambda^2s^2},1\big]\big)\,.
$$
As a consequence,
\begin{align*}
G(s,\cdot)^{-1}(\aleph)&
\;=\; 
H_{\pm}^s\big(\aleph\cap\big\{\sqrt{1-\lambda^2s^2}\big\}\big)
\;\cup\;
\bigcup\limits_{\sigma=\pm}H_{\sigma}^s\big(\aleph\cap\big(\sqrt{1-\lambda^2s^2},1\big]\big)\\
&
\;=\;
\left(H_{\pm}^s(\aleph)\cap\{-\lambda s\}\right)
\;\cup\;
\bigcup\limits_{\sigma=\pm}H_{\sigma}^s\big(\aleph\cap\big(\sqrt{1-\lambda^2s^2},1\big]\big)
\;.
\end{align*}
Now $H_{\pm}^s$ are locally Lipschitz continuous on $\big(\sqrt{1-\lambda^2s^2},1\big]$.
This implies that also $G(s,\cdot)^{-1}(\aleph)$ is an $\textnormal{\textbf{x}}$-nullset. Due to the absolutely continuous distribution of $Y$ w.r.t.~$\textbf{x}$, one therefore has
\begin{align*}
\mathbb{P}\left(G(r,Y)\in\aleph\right)&=\int\textnormal{d}\mathbb{P}(r\in\cdot)(s)\textnormal{ }\mathbb{P}\left(G(s,Y)\in\aleph\right)\\
&
\;=\; \int_{(0,1]}\textnormal{d}\mathbb{P}(r\in\cdot)(s)\textnormal{ }\mathbb{P}\left(Y\in G(s,\cdot)^{-1}(\aleph)\right)+\chi_{\aleph}(1)\textnormal{ }\mathbb{P}(r=0)\\
&
\;=\;
\chi_{\aleph}(1)\textnormal{ }\mathbb{P}(r=0)\\
&
\;=\;
\mathbb{P}(r=0)\textnormal{ }\delta_1(\aleph)
\;,
\end{align*} 
and this concludes the proof. 
\hfill $\Box$

\vspace{.2cm}

\noindent {\bf Proof} of Lemma~\ref{lem-2}.
Let $(s,\mathscr{U})\in [0,1]\times\textnormal{O}(\mathsf{L}+1)$. Then,
\begin{align*}
\begin{split}
(\mathbf{1}+\lambda s\mathscr{U})\cdot v 
\; &=\;\left\|(\mathbf{1}+\lambda s\mathscr{U})v\right\|^{-1} (\mathbf{1}+\lambda s\mathscr{U})v
\\
&=\;
\left\|\mathscr{V}(\mathbf{1}+\lambda s\mathscr{U})v\right\|^{-1}\mathscr{V}^*\mathscr{V}(\mathbf{1}+\lambda s\mathscr{U})v
\\
&=\;
\left\|(\mathbf{1}+\lambda s\mathscr{V}\mathscr{U})v\right\|^{-1}
\mathscr{V}^*(\mathbf{1}+\lambda s\mathscr{V}\mathscr{U})v
\\
&=\;\mathscr{V}^*(\mathbf{1}+\lambda s\mathscr{V}\mathscr{U})\cdot v
\end{split}
\end{align*}
holds. But $\mathscr{V}U$ is distributed identically to $U$ and this implies \eqref{ac_distribution_8}. As $(\mathbf{1}+\lambda s\mathscr{U})\cdot Vv= V(\mathbf{1}+\lambda sV^*\mathscr{U}V)\cdot v$, the proof of \eqref{ac_distribution_8first} follows in a similar manner.
\hfill $\Box$

\vspace{.2cm}

\noindent {\bf Proof} of Lemma~\ref{lem-2a}.
By Lemma~\ref{lem-1}, the pure point part of the probability distribution $\varpi_{r,\lambda}$ of the \mbox{\footnotesize $\big[\sqrt{1-\lambda^2},1\big]$}-valued random variable $Z=\left\langle v,(\mathbf{1}+\lambda rU)\cdot v\right\rangle$ is given by $\varpi_{r,\lambda}^{\textnormal{pp}}=\mathbb{P}(r=0)\textnormal{ }\delta_1$. This implies the first equality in \eqref{lem-2a-ppsc}, since $Z=1$ is equivalent to $(\mathbf{1}+\lambda rU)\cdot v=v$.

\vspace{.1cm}

As for the continuous part of $\varrho_{r,\lambda,v}$, let us write
\begin{equation}
\label{eq-Zvperp}
(\mathbf{1}+\lambda rU)\cdot v
\;=\;Z\,v\;+\;(1-Z^2)^{\frac{1}{2}}\,v_\perp
\;,
\end{equation}
where $v_\perp\in\mathbb{S}^{\mathsf{L}}$ is a random unit vector orthogonal to $v$. By Lemma~\ref{lem-2}, the distribution of $v_\perp$ is invariant under the fixed point group of $v$, namely the action of \mbox{$\{\mathscr{V}\in \textnormal{O}(\mathsf{L}+1)\,:\,\mathscr{V}v=v\}$}. Thus the distribution of $v_\perp$ is given by the push-forward of $(i_v)_*(\nu_{\mathsf{L}-1})$ under a natural embedding $i_v:\SM^{\mathsf{L}-1}\to\{w\in \SM^{\mathsf{L}}\,:\,w\perp v\}$. Furthermore, $Z$ and $v_\perp$ are independent. Indeed, by~\eqref{eq-ActionFormula} $Z$ only depends on the component $Y=\langle v,Uv\rangle$ of the vector $Uv$ in the direction of~$v$, while for $Z\not=1$
$$
v_\perp\;=\;
\frac{P_\perp\,\big((\mathbf{1}+\lambda rU)\cdot v\big)}{(1-Z^2)^{\frac{1}{2}}}
\;=\;
\frac{P_\perp Uv}{\|P_\perp Uv\|}
\;,
$$
with $P_\perp$ being the projection onto the orthogonal complement of the span of $v$, so that $v_\perp$ only depends on the direction of the component of $Uv$ orthogonal to $v$, which is independent of the component parallel to $v$.  

\vspace{.1cm}

Now by the above and Lemma~\ref{lem-1} the distribution of $(Z,(i_v)^{-1}(v_\perp))$ is equal to $(\varpi_{r,\lambda}^{\textnormal{pp}}+\varpi_{r,\lambda}^{\textnormal{ac}})\otimes \nu_{\mathsf{L}-1}$ and therefore 
\begin{align}
\label{lem-2a-eq-3}
\varrho_{r,\lambda,v}
\;=\;
(F_v)_*\big(\varpi_{r,\lambda}^{\textnormal{pp}}\otimes \nu_{\mathsf{L}-1}\big)
\,+\,
(F_v)_*\big(\varpi_{r,\lambda}^{\textnormal{ac}}\otimes \nu_{\mathsf{L}-1}\big)
\;,
\end{align}
where the function
$$
F_v\,:\,\big[\sqrt{1-\lambda^2},1\big]\times\mathbb{S}^{\mathsf{L}-1}\,\to\, \mathbb{S}^{\mathsf{L}}\;,
\qquad (z,w)\,\mapsto \,z\,v\;+\;(1-z^2)^{\frac{1}{2}}\,i_v(w)
$$
maps the set $\{1\}\times \mathbb{S}^{\mathsf{L}-1}$ to the point $v$ and the set $\mbox{\footnotesize$\big[\sqrt{1-\lambda^2},1\big)$}\times\mathbb{S}^{\mathsf{L}-1}$ bijectively onto $\left\{u\in\mathbb{S}^{\mathsf{L}}:\langle u,v\rangle\in\mbox{\footnotesize$\big[\sqrt{1-\lambda^2},1\big)$}\right\}$. Using
$$
(F_v)_*\big(\varpi_{r,\lambda}^{\textnormal{pp}}\otimes \nu_{\mathsf{L}-1}\big)(\{v\})
\;=\;
\mathbb{P}(r=0)
$$
and
\begin{align*}
(F_v)_*\big(\varpi_{r,\lambda}^{\textnormal{pp}}\otimes \nu_{\mathsf{L}-1}\big)(\mathbb{S}^{\mathsf{L}}\setminus\{v\})
&
\;=\;
\big(\varpi_{r,\lambda}^{\textnormal{pp}}\otimes \nu_{\mathsf{L}-1}\big)\big(\mbox{\footnotesize $\big[\sqrt{1-\lambda^2},1\big)$}\times \mathbb{S}^{\mathsf{L}-1}\big)\\
&
\;=\;
\varpi_{r,\lambda}^{\textnormal{pp}}\big(\mbox{\footnotesize $\big[\sqrt{1-\lambda^2},1\big)$}\big)\textnormal{ }\nu_{\mathsf{L}-1}\big(\mathbb{S}^{\mathsf{L}-1}\big)\\
&
\;=\;
0
\;,
\end{align*}
combined with the first identity in \eqref{lem-2a-ppsc}, one infers that $(F_v)_*\big(\varpi_{r,\lambda}^{\textnormal{pp}}\otimes \nu_{\mathsf{L}-1}\big)$ and $\varrho_{r,\lambda,v}^{\textnormal{pp}}$ coincide. This, in turn, implies together with~\eqref{lem-2a-eq-3} that
\begin{align}
\label{lem-2a-eq-4}
(F_v)_*\big(\varpi_{r,\lambda}^{\textnormal{ac}}\otimes \nu_{\mathsf{L}-1}\big)
\;=\;
\varrho_{r,\lambda,v}^{\textnormal{ac}}
\,+\,
\varrho_{r,\lambda,v}^{\textnormal{sc}}
\end{align}
holds and~\eqref{lem-2a-eq-4} is continuous.
Now, since the restriction of $F_v$ to any compact subset of $\mbox{\footnotesize$\big[\sqrt{1-\lambda^2},1\big)$}\times\mathbb{S}^{\mathsf{L}-1}$ is bi-Lipschitz, the preimage of any $\nu_{\mathsf{L}}$-nullset contained in $\mathbb{S}^{\mathsf{L}}\setminus\{v\}$ under $F_v$ is an $\textbf{x}\otimes \nu_{\mathsf{L}-1}$-nullset and hence, in particular, a $\varpi_{r,\lambda}^{\textnormal{ac}}\otimes \nu_{\mathsf{L}-1}$-nullset. Therefore,~\eqref{lem-2a-eq-4} is even absoluely continuous, {\it i.e.}, the second identity in \eqref{lem-2a-ppsc} holds.

\vspace{.1cm}

As for the proof of~\eqref{lem-2a-ac}, one may assume that $v\neq w$, as~\eqref{lem-2a-ac} is trivial otherwise. Now let $V\in \textnormal{O}(\mathsf{L}+1)$ such that $Vv=w$. Since $\langle v,Vv\rangle=\langle v,V^*v\rangle$, the vectors $Vv$ and $V^*v$ have the same projection in the direction of $v$. Hence there exists a $\mathscr{V}\in \textnormal{O}(\mathsf{L}+1)$ satisfying $\mathscr{V}v=v$ such that $\mathscr{V}V^*v=Vv$. Now by applying both \eqref{ac_distribution_8first} and \eqref{ac_distribution_8} one deduces that for every ball $B_\epsilon(v)\subset \mathbb{S}^{\mathsf{L}}$ of radius $\epsilon>0$ around $v$
\begin{align*}
\varrho_{r,\lambda,w}(B_\epsilon(v))
&
\;=\;
\mathbb{P}\left((\mathbf{1}+\lambda rU)\cdot w\in B_\epsilon(v)\right)
\\
&
\;=\;
\mathbb{P}\left((\mathbf{1}+\lambda rU)\cdot v\in \mathscr{V}V^*B_\epsilon(v) \right)
\\
&
\;=\;
\mathbb{P}\left((\mathbf{1}+\lambda rU)\cdot v\in B_\epsilon(w)\right)
\\
&
\;=\;
\varrho_{r,\lambda,v}(B_\epsilon(w))
\,.
\end{align*}
If $\epsilon<\|v-w\|$ and due to \eqref{lem-2a-ppsc}, this is equivalent to
$$
\varrho_{r,\lambda,w}^{\textnormal{ac}}(B_\epsilon(v))
\;=\;
\varrho_{r,\lambda,v}^{\textnormal{ac}}(B_\epsilon(w))
\;.
$$
Taking the Radon-Nikodym derivatives now implies~\eqref{lem-2a-ac}.
\hfill $\Box$

\vspace{.2cm}

\noindent {\bf Proof} of Lemma~\ref{lem-4}. The map $\mathcal{R}^{-1}\cdot$ is Lipschitz because for all $v_1,v_2\in\mathbb{S}^{\mathsf{L}}$ one has
\begin{align*}
\left\|\mathcal{R}^{-1}\cdot v_1-\mathcal{R}^{-1}\cdot v_2\right\|
&
\;=\;
\frac{1}{\left\|\mathcal{R}^{-1}v_2\right\|}\,
\left\|\Big(\left\|\mathcal{R}^{-1}v_2\right\|-\left\|\mathcal{R}^{-1}v_1\right\|\Big)\frac{\mathcal{R}^{-1}v_1}{\left\|\mathcal{R}^{-1}v_1\right\|}+\mathcal{R}^{-1}(v_1-v_2)\right\|
\\
&
\;\leq\;
\left\|\mathcal{R}\right\|\,
\left(\Big|\left\|\mathcal{R}^{-1}v_2\right\|-\left\|\mathcal{R}^{-1}v_1\right\|\Big|+\left\|\mathcal{R}^{-1}(v_1-v_2)\right\|\right)
\\
&
\;\leq \;
2\,\left\|\mathcal{R}\right\|\,
\left\|\mathcal{R}^{-1}(v_1-v_2)\right\|
\\
&
\;\leq\; 2\,\left\|\mathcal{R}\right\|\,\left\|\mathcal{R}^{-1}\right\|\left\|v_1-v_2\right\|
\;.
\end{align*}
%
Thus $\mathcal{R}^{-1}\cdot \aleph$ is a $\nu_{\mathsf{L}}$-nullset for any $\nu_{\mathsf{L}}$-nullset $\aleph$, which implies the claim.
\hfill $\Box$

\vspace{.2cm}

\noindent \textbf{Proof} of Lemma~\ref{large-microscopic-gap}. Let $\lambda<2^{-4}{\delta'\mathcal{R}_i}$ where ${\delta'\mathcal{R}_i}=\min\{\delta\mathcal{R}_i,\mbox{\small $\frac{1}{2}$}\}$. Let us denote the orthogonal projections onto $\mathbb{R}^{\mathsf{L}+1-i}\times\{0\}^{i}$ and $\{0\}^{\mathsf{L}+1-i}\times\mathbb{R}^{i}$ by $\mathscr{P}_i^{\uparrow}$ and $\mathscr{P}_i^{\downarrow}$, respectively. One has the estimates
\begin{align*}
\|\mathscr{P}_i^{\uparrow}(\mathcal{R}\cdot w)\|^2
&
\;=\;\left(1+\|\mathscr{P}^{\downarrow}_i\mathcal{R}w\|^2\|\mathscr{P}^{\uparrow}_i\mathcal{R}w\|^{-2}\right)^{-1}\\&\leq\left(1+\kappa_i^2\kappa_{i+1}^{-2}\|\mathscr{P}^{\downarrow}_iw\|^2\|\mathscr{P}^{\uparrow}_iw\|^{-2}\right)^{-1}\\
&
\;=\;
\|\mathscr{P}^{\uparrow}_iw\|^2\left(1+\left(\kappa_i\kappa_{i+1}^{-1}+1\right)\delta\mathcal{R}_i\|\mathscr{P}^{\downarrow}_iw\|^2\right)^{-1}\\
&
\;\leq\; \|\mathscr{P}^{\uparrow}_iw\|^2\left(1+2\delta\mathcal{R}_i\|\mathscr{P}^{\downarrow}_iw\|^2\right)^{-1}\\
&
\;\leq\;  \|\mathscr{P}^{\uparrow}_iw\|^2\left(1-{\delta'\mathcal{R}_i}\|\mathscr{P}^{\downarrow}_iw\|^2\right)\,,
\end{align*}
and
\begin{align*}
\|\mathscr{P}^{\uparrow}_i((\mathbf{1}+\lambda s\mathscr{U})\cdot w)\|^2
&
\;=\;
1\,-\,\|\mathscr{P}^{\downarrow}_i(\mathbf{1}+\lambda s\mathscr{U}) w\|^2\|(\mathbf{1}+\lambda s\mathscr{U})w\|^{-2}\\
&
\;\leq\; 
1\,-\,\|\mathscr{P}^{\downarrow}_i(\mathbf{1}+\lambda s\mathscr{U}) w\|^2\left[2-\|(\mathbf{1}+\lambda s\mathscr{U})w\|^{2}\right]\\
&
\;=\;
\|\mathscr{P}^{\uparrow}_iw\|^2\,+\,2\lambda s\left\langle \left(\|\mathscr{P}^{\downarrow}_i(\mathbf{1}+\lambda s\mathscr{U})w\|^2-\mathscr{P}^{\downarrow}_i\right)w,\mathscr{U}w\right\rangle
\\
&\qquad\;+\;\lambda^2s^2\left\langle \mathscr{P}^{\downarrow}_iw,\left(\mathscr{P}^{\downarrow}_i+2\lambda s \mathscr{U}\right)w\right\rangle\\
&
\;\leq \;
\|\mathscr{P}^{\uparrow}_iw\|^2\,+\,2\lambda s(1+\lambda s)^2\,+\,\lambda^2s^2(1+2\lambda s)\\
&
\;\leq\;
\|\mathscr{P}^{\uparrow}_iw\|^2\,+\,\frac{7}{2}\lambda
\end{align*}
for all $w\in\mathbb{S}^{\mathsf{L}}$, $s\in [0,1]$ and $\mathscr{U}\in\textnormal{O}(\mathsf{L}+1)$. Combining these estimates leads to
\begin{align}
\label{large-microscopic-gap-3}
\|\mathscr{P}_i^{\uparrow}((\mathbf{1}+\lambda s\mathscr{U})\mathcal{R}\cdot v)\|^2
\,-\,
(1-\lambda/2)\|\mathscr{P}^{\uparrow}_iv\|^2
\;\leq\; 
{\delta'\mathcal{R}_i}\|\mathscr{P}^{\uparrow}_iv\|^4-{\delta'\mathcal{R}_i}\|\mathscr{P}^{\uparrow}_iv\|^2+4\lambda
\;,
\end{align}
holding for all $v\in\mathbb{S}^{\mathsf{L}}$, $s\in [0,1]$ and $\mathscr{U}\in\textnormal{O}(\mathsf{L}+1)$. For these  parameters, \eqref{large-microscopic-gap-3} now implies the following statements:
\begin{align*}
& 
\mbox{\rm (i) }\;\mbox{\small $\|\mathscr{P}^{\uparrow}_iv\|^2\in\mbox{\footnotesize $\frac{1}{2}$}\left(1+
[-1,1]\mbox{\footnotesize $\sqrt{1-16\lambda({\delta'\mathcal{R}_i})^{-1}}$}\hspace{0.3mm}\right)\textnormal{ }\Longrightarrow\textnormal{ }
\|\mathscr{P}_i^{\uparrow}((\mathbf{1}+\lambda s\mathscr{U})\mathcal{R}\cdot v)\|^2\,\leq\,(1-\tfrac{\lambda}{2})\|\mathscr{P}^{\uparrow}_iv\|^2
\,,$}
\\
&
\mbox{\rm (ii) }\mbox{\small $\|\mathscr{P}^{\uparrow}_iv\|^2< \frac{1}{2}\left(1-\mbox{\footnotesize $\sqrt{1-16\lambda({\delta'\mathcal{R}_i})^{-1}}$}\hspace{0.3mm}\right)\;\Longrightarrow\textnormal{ }\|\mathscr{P}_i^{\uparrow}((\mathbf{1}+\lambda s\mathscr{U})\mathcal{R}\cdot v)\|^2\,<\, \frac{1}{2}\big(1-\mbox{\footnotesize $\sqrt{1-16\lambda(\delta'\mathcal{R}_i)^{-1}}$}\big)$}.
\end{align*}
As for the dynamics $\{v_n\}_{n\in\mathbb{N}}$ defined by~\eqref{dynamics_sphere}~and~\eqref{random_matrix}, statements (i) and (ii) guarantee the existence of $N\in\mathbb{N}$ such that all $n\in\mathbb{N}$ satisfy
$$
\mathbb{P}\left(\|\mathscr{P}^{\uparrow}_iv_{n+N}\|^2<\mbox{\small $ \frac{1}{2}\left(1-\mbox{\footnotesize $\sqrt{1-16\lambda({\delta'\mathcal{R}_i})^{-1}}$}\hspace{0.3mm}\right)$}\textnormal{ }\Big|\textnormal{ }\|\mathscr{P}^{\uparrow}_iv_n\|^2\leq\mbox{\small $ \frac{1}{2}\left(1+\mbox{\footnotesize $\sqrt{1-16\lambda({\delta'\mathcal{R}_i})^{-1}}$}\hspace{0.3mm}\right)$}\right)
\;=\;1
\;,
$$
and thus
$$
\mathbb{P}\left(\|\mathscr{P}^{\uparrow}_iv_{n+N}\|^2<\mbox{\small $ \frac{1}{2}\left(1-\mbox{\footnotesize $\sqrt{1-16\lambda({\delta'\mathcal{R}_i})^{-1}}$}\hspace{0.3mm}\right)$}\right)
\;\geq\;
\mathbb{P}\left(\|\mathscr{P}^{\uparrow}_iv_n\|^2\leq\mbox{\small $ \frac{1}{2}\left(1+\mbox{\footnotesize $\sqrt{1-16\lambda({\delta'\mathcal{R}_i})^{-1}}$}\hspace{0.3mm}\right)$}\right)\,.
$$
Therefore,
\begin{align*}
&\mu_{r,\lambda}\left(\left\{v\in\mathbb{S}^{\mathsf{L}}: \|\mathscr{P}^{\uparrow}_iv\|^2<\mbox{\small $ \frac{1}{2}\left(1-\mbox{\footnotesize $\sqrt{1-16\lambda({\delta'\mathcal{R}_i})^{-1}}$}\hspace{0.3mm}\right)$}\right\}\right)\\
&\qquad
\;=\;
\int\limits_{\mathbb{S}^{\mathsf{L}}}\textnormal{d}\mu_{r,\lambda}(v_0)\textnormal{ }\mathbb{P}\left(\|\mathscr{P}^{\uparrow}_iv_{N}\|^2<\mbox{\small $ \frac{1}{2}\left(1-\mbox{\footnotesize $\sqrt{1-16\lambda({\delta'\mathcal{R}_i})^{-1}}$}\hspace{0.3mm}\right)$}\right)\\
&
\qquad
\;\geq\;
\int\limits_{\mathbb{S}^{\mathsf{L}}}\textnormal{d}\mu_{r,\lambda}(v_0)\textnormal{ }\chi_{\left\{v\in\mathbb{S}^{\mathsf{L}}: \|\mathscr{P}^{\uparrow}_iv\|^2\leq\mbox{\small $ \frac{1}{2}\left(1+\mbox{\footnotesize $\sqrt{1-16\lambda({\delta'\mathcal{R}_i})^{-1}}$}\hspace{0.3mm}\right)$}\right\}}(v_0)\\
&
\qquad
\;=\;
\mu_{r,\lambda}\left(\left\{v\in\mathbb{S}^{\mathsf{L}}: \|\mathscr{P}^{\uparrow}_iv\|^2\leq\mbox{\small $ \frac{1}{2}\left(1+\mbox{\footnotesize $\sqrt{1-16\lambda({\delta'\mathcal{R}_i})^{-1}}$}\hspace{0.3mm}\right)$}\right\}\right)\,,
\end{align*}
which is equivalent to
$$
\mu_{r,\lambda}\left(\left\{v\in\mathbb{S}^{\mathsf{L}}: \mbox{\small $ \frac{1}{2}\left(1-\mbox{\footnotesize $\sqrt{1-16\lambda({\delta'\mathcal{R}_i})^{-1}}$}\hspace{0.3mm}\right)$}<\|\mathscr{P}^{\uparrow}_iv\|^2\leq \mbox{\small $ \frac{1}{2}\left(1+\mbox{\footnotesize $\sqrt{1-16\lambda({\delta'\mathcal{R}_i})^{-1}}$}\hspace{0.3mm}\right)$} \right\}\right)
\;=\;
0\,,
$$
which proves $\textnormal{supp}(\mu_{r,\lambda})\neq \mathbb{S}^{\mathsf{L}}$. 
\hfill $\Box$

\vspace{.2cm}

\noindent {\bf Proof} of Lemma~\ref{reach_point}. The aim is to construct $(s_n)_{n=1,\ldots,N}$ in $\textnormal{supp}(r)$ and $(\mathscr{U}_n)_{n=1,\ldots,N}$ in $\textnormal{O}(\mathsf{L}+1)$ such that for a given couple $u,w\in \mathbb{S}^{\mathsf{L}}$ 
$$
\prod_{n=1}^{N}\mathcal{R}\left(\mathbf{1}+\lambda s_n\mathscr{U}_n\right)\cdot u
\;=\;w\;,
$$
where $u$ is an initial condition which we may choose to be the stable fixed point $e_{\mathsf{L}+1}$, as the motion from some arbitrary $u$ towards this stable fixed point $e_{\mathsf{L}+1}$ via a finite path is somewhat straightforward and is thus left to the reader. To accommodate notations, let us use the unit vectors $\tilde{e}_j=e_{\mathsf{L}+2-j}$ so that $\Rr \tilde{e}_j=\kappa_j\tilde{e}_j$. Then $w=\sum_{j=1}^{\mathsf{L}+1}w_j \tilde{e}_j=(w_1,\ldots, w_{\mathsf{L}+1})^{\intercal}$. Further let us introduce $\mathsf{K}=\max\left\{\mathsf{J}\in\{1,\dots,\mathsf{L}+1\}: w_{\mathsf{J}}\neq 0\right\}$. 

\vspace{.2cm}

\noindent {{\bf Step 1.}} \textit{There exist $N_1\in\mathbb{N}_0$ and $(\mathscr{U}_n^{\pm})_{n=1,\ldots,N_1}$ in $\mbox{\rm O}(\mathsf{L}+1)$  such that}
$$
\prod_{n=1}^{N_1}\mathcal{R}\left(\mathbf{1}+\lambda \mathscr{U}_n^{\pm}\right)\cdot \tilde{e}_1
\;=\;
\pm \,\tilde{e}_{\mathsf{K}}
\;.
$$

\vspace{.1cm}

\noindent One can assume $\mathsf{K}\neq 1$ as the statement is trivial otherwise. Let us set
$$
\mathscr{U}_1^{\pm}
\;=\;
\begin{pmatrix}
1\\ & \ddots\\&& 1 \\ &&&0&\pm 1\\&&&1&0
\end{pmatrix}
\;.
$$
Then $\mathcal{R}\left(\mathbf{1}+\lambda\mathscr{U}_1^{\pm}\right)\tilde{e}_1=\pm\kappa_2\lambda \tilde{e}_2+\kappa_1 \tilde{e}_1\,$. Next for $n\in\{2,\dots,N_1-1\}$ with $N_1$ to be chosen later, we choose $\mathscr{U}_n^{\pm}=\textnormal{diag}(1,\dots,1,-1)$. It follows that $\mathcal{R}\left(\mathbf{1}+\lambda\mathscr{U}_n^{\pm}\right)=\textnormal{diag}(\kappa_{\mathsf{L}+1}(1+\lambda),\dots,\kappa_2(1+\lambda),\kappa_1(1-\lambda))\,$ so that
$$
\prod_{n=1}^{N_1-1}\mathcal{R}\left(\mathbf{1}+\lambda \mathscr{U}_n^{\pm}\right)\tilde{e}_1
\;=\;
\pm\left[\kappa_2(1+\lambda)\right]^{N_1-1}\kappa_2\lambda \tilde{e}_2+\left[\kappa_1(1-\lambda)\right]^{N_1-1}\kappa_1 \tilde{e}_1
\;.
$$
The assumption on $\lambda$ guarantees that $\kappa_2(1+\lambda)>\kappa_1(1-\lambda)$ and therefore one can choose $N_1$ such that 
$$
\left[\kappa_1(1-\lambda)\right]^{N_1-1}\kappa_1
\;\leq \;
\lambda\left(\left[\kappa_2(1+\lambda)\right]^{N_1-1}\kappa_2\lambda\right)
\;.
$$
Hence, there exists some $\epsilon\leq \lambda$ such that the proportionality relation
$$
\prod_{n=1}^{N_1-1}\mathcal{R}\left(\mathbf{1}+\lambda \mathscr{U}_n^{\pm}\right)\tilde{e}_1
\;\propto\;
\pm \,\tilde{e}_2\,+\,\epsilon \tilde{e}_1
$$
holds. Now, one can choose $\mathscr{U}_{N_1}^{\pm}$ in such a way that 
$$
\left\langle\lambda\mathscr{U}_{N_1}^{\pm} (\pm \tilde{e}_2+\epsilon \tilde{e}_1), \tilde{e}_1\right\rangle 
\;=\;-\,\epsilon
$$
and 
$$
\left\langle\lambda\mathscr{U}_{N_1}^{\pm} (\pm \tilde{e}_2+\epsilon \tilde{e}_1), \tilde{e}_{\mathsf{J}}\right\rangle 
\;=\; 0\;, 
\qquad 
\forall\;\;\mathsf{J}\in\{3,\dots,\mathsf{L}+1\}
\;,
$$
are satisfied. It follows that
$$
\left(\mathbf{1}+\lambda \mathscr{U}_{N_1}^{\pm}\right)\prod_{n=1}^{N_1-1}\mathcal{R}\left(\mathbf{1}+\lambda \mathscr{U}_n^{\pm}\right)\tilde{e}_1
\;\propto\;
\pm \,\tilde{e}_2
\;,
$$
and thus
$$
\prod_{n=1}^{N_1}\mathcal{R}\left(\mathbf{1}+\lambda \mathscr{U}_n^{\pm}\right)\cdot \tilde{e}_1
\;=\;
\pm \,\tilde{e}_2
$$
holds. In the same vein, one may construct paths from $\tilde{e}_{\mathsf{J-1}}$ to $\tilde{e}_{\mathsf{J}}$ for $\mathsf{J}\in\{3,\dots, \mathsf{K}\}$. This finishes the proof of Step 1.

\vspace{.1cm}

Next let us set $\tilde{\mathsf{K}}=\min\left\{\mathsf{J}\in\{1,\dots,\mathsf{K}\}: \kappa_{\mathsf{J}}=\kappa_{\mathsf{K}}\right\}\,.$

\vspace{.1cm}

\noindent {{\bf Step 2.}} \textit{There exist $N_2\in\mathbb{N}_0$, sequences $(s_n)_{n=1,\ldots,N_2}$ in $ \textnormal{supp}(r)$ and $(\mathscr{U}_n)_{n=1,\ldots,N_2}$ in $\mbox{\rm O}(\mathsf{L}+1)$ such that}
\begin{align}
\label{reach_point_0}
\prod_{n=1}^{N_2}\mathcal{R}\left(\mathbf{1}+\lambda s_n\mathscr{U}_n\right)\cdot \tilde{e}_{\mathsf{K}}
\;=\; 
\sum_{{\mathsf{J}}=\tilde{\mathsf{K}}}^{\mathsf{K}}w_{\mathsf{J}}\tilde{e}_{\mathsf{J}}
\Big\|\sum_{{\mathsf{J}}=\tilde{\mathsf{K}}}^{\mathsf{K}}w_{\mathsf{J}}\tilde{e}_{\mathsf{J}}
\Big\|^{-1}
\;.
\end{align}

\vspace{.1cm}

\noindent Let $\tilde{U}_{\mathsf{K}-\tilde{\mathsf{K}}+1}$ be an $\textnormal{O}(\mathsf{K}-\tilde{\mathsf{K}}+1)$-valued random variable distributed according to the Haar measure. It induces an $\textnormal{O}(\mathsf{L}+1)$-valued random variable by $\tilde{U}_{\tilde{\mathsf{K}},\mathsf{K}}= \mathbf{1}_{\mathsf{L}+1-\mathsf{K}}\oplus \tilde{U}_{\mathsf{K}-\tilde{\mathsf{K}}+1}\oplus \mathbf{1}_{\tilde{\mathsf{K}}-1}$. Since $\kappa_{\mathsf{K}}=\dots=\kappa_{\tilde{\mathsf{K}}}$, the action $\mathcal{R}\cdot$ is trivial on the submanifold $\mathscr{S}_{\tilde{\mathsf{K}},\mathsf{K}}=\{0\}^{\mathsf{L}+1-\mathsf{K}}\times \mathbb{S}^{\mathsf{K}-\tilde{\mathsf{K}}}\times \{0\}^{\tilde{\mathsf{K}}-1}$  and commutes with $(\mathbf{1}+\lambda r\tilde{U}_{\tilde{\mathsf{K}},\mathsf{K}})\cdot $ which acts transitively on $\mathscr{S}_{\tilde{\mathsf{K}},\mathsf{K}}$ (see  Proposition~\ref{prop-RPP} for a detailed proof). This shows Step 2. Combined with the above, the next step concludes the proof.

\vspace{.1cm}

\noindent {{\bf Step 3.}} \textit{There exist $N_3\in\mathbb{N}_0$ and $(\mathscr{U}_n)_{n=1,\ldots,N_3}$ in $\mbox{\rm O}(\mathsf{L}+1)$ such that}
\begin{align}\label{reach_point_2}
\prod_{n=1}^{N_3}\mathcal{R}\left(\mathbf{1}+\lambda \mathscr{U}_n\right)\cdot \sum_{{\mathsf{J}}=\tilde{\mathsf{K}}}^{\mathsf{K}}w_{\mathsf{J}}\tilde{e}_{\mathsf{J}}
\Big\|\sum_{{\mathsf{J}}=\tilde{\mathsf{K}}}^{\mathsf{K}}w_{\mathsf{J}}\tilde{e}_{\mathsf{J}}\Big\|^{-1}
\;=\; w
\;.
\end{align}

\vspace{.1cm}

\noindent One can assume $\left(w_{\tilde{\mathsf{K}}-1},\dots, w_1\right)^{\intercal}\neq 0$, as the statement is trivial otherwise. Let us abbreviate
$$
y
\;=\;
(w_{\mathsf{K}},\dots, w_{\tilde{\mathsf{K}}})^{\intercal}
\Big\|\sum_{{\mathsf{J}}=\tilde{\mathsf{K}}}^{\mathsf{K}}w_{\mathsf{J}}\tilde{e}_{\mathsf{J}}\Big\|^{-1}
\;\in\; \mathbb{R}^{\mathsf{K}-\tilde{\mathsf{K}}+1}
$$
and use the notation $(x_{\mathsf{L}+1},\dots,x_1)^{\intercal}:=\mathscr{U}_1(0,\dots,0,y,0,\dots 0)^{\intercal}$ with $\mathscr{U}_1$ to be chosen later. Set $\mathscr{U}_n=\mathbf{1}$ for $n\in\{2,\dots, N_3\}$,  where $N_3$ will also be chosen later. Now the l.h.s. of~\eqref{reach_point_2} is proportional to
$$
\mathcal{R}^{N_3}\left(\mathbf{1}+\lambda \mathscr{U}_1\right)(0,\dots,0,y,0,\dots 0)^{\intercal}
\;=\;
\begin{pmatrix}\lambda(\kappa_{\mathsf{L}+1}^{N_3} x_{\mathsf{L}+1},\dots, \kappa_{\mathsf{K}+1}^{N_3} x_{\mathsf{K}+1})^{\intercal} \\ \kappa_{\mathsf{K}}^{N_3}\left(y+\lambda (x_{\mathsf{K}},\dots,x_{\tilde{\mathsf{K}}})^{\intercal}\right)\\ \lambda(\kappa_{\tilde{\mathsf{K}}-1}^{N_3} x_{\tilde{\mathsf{K}}-1},\dots,\kappa^{N_3}_1x_1)^{\intercal}\end{pmatrix}
\;,
$$
which in turn has to be proportional to $w$ so that, for some $c\in (0,\infty)$,
\begin{align}
\label{reach_point_4}
\begin{pmatrix}\lambda(\kappa_{\mathsf{L}+1}^{N_3} x_{\mathsf{L}+1},\dots, \kappa_{\mathsf{K}+1}^{N_3} x_{\mathsf{K}+1})^{\intercal} \\ \kappa_{\mathsf{K}}^{N_3}\left(y+\lambda (x_{\mathsf{K}},\dots,x_{\tilde{\mathsf{K}}})^{\intercal}\right)\\ \lambda(\kappa_{\tilde{\mathsf{K}}-1}^{N_3} x_{\tilde{\mathsf{K}}-1},\dots,\kappa^{N_3}_1x_1)^{\intercal}\end{pmatrix}
\;=\;
c\,
\begin{pmatrix}(w_{\mathsf{L}+1},\dots,w_{\mathsf{K}+1})^{\intercal}\\(w_{\mathsf{K}},\dots, w_{\tilde{\mathsf{K}}})^{\intercal}\\(w_{\tilde{\mathsf{K}}-1},\dots,w_1)^{\intercal}\end{pmatrix}
\;.
\end{align}
Now $(w_{\mathsf{L}+1},\dots,w_{\mathsf{K}+1})^{\intercal}=(0,\dots,0)^{\intercal}$ requires the choice $(x_{\mathsf{L}+1},\dots,x_{\mathsf{K}+1})^{\intercal}=(0,\dots,0)^{\intercal}$. Moreover, since $y$ is proportional to $(w_{\mathsf{K}},\dots, w_{\tilde{\mathsf{K}}})^{\intercal}$, the middle part of~\eqref{reach_point_4} forces us to set
\begin{align*}
(x_{\mathsf{K}},\dots, x_{\tilde{\mathsf{K}}})^{\intercal}
\;=\;
y\big(1-x_1^2-\dots-x_{\tilde{\mathsf{K}}-1}^2\big)^{\frac{1}{2}}
\;,
\end{align*}
where $x_{\tilde{\mathsf{K}}-1},\dots,x_1$ are given by the lower part of~\eqref{reach_point_4} as
\begin{align*}
x_{\tilde{\mathsf{K}}-1}\;=\;\frac{c}{\lambda}\,\frac{w_{\tilde{\mathsf{K}}-1}}{\kappa_{\tilde{\mathsf{K}}-1}^{N_3}}\;,\quad
\dots\;,\quad
x_{1}\;=\;\frac{c}{\lambda}\,\frac{w_{1}}{\kappa_{1}^{N_3}}\;,
\end{align*}
where $c$ and $N_3$ have still to be chosen appropriately in order to satisfy the remaining middle part, which is now of the (scalar) form
$$
\kappa^{N_3}_{\mathsf{K}}
\Big[
1+\lambda
\Big( 1-\frac{c^2}{\lambda^2}
\sum_{{\mathsf{J}}=1}^{\tilde{\mathsf{K}}-1}
\Big(
\frac{w_{{\mathsf{J}}}}{\kappa_{{\mathsf{J}}}^{N_3}}\Big)^2 
\Big)^{\frac{1}{2}} 
\Big]
\;=\;
c\,
\left\|(w_{\mathsf{K}},\dots, w_{\tilde{\mathsf{K}}})^{\intercal}\right\|
\;.
$$
It hence suffices to demonstrate the existence of some $N_3\in\mathbb{N}$ such that the function
$$
c\in(0,\infty)
\;\mapsto\;
f_{N_3}(c)
\;=\;
c\left\|(w_{\mathsf{K}},\dots, w_{\tilde{\mathsf{K}}})^{\intercal}\right\|
\;-\;
\kappa^{N_3}_{\mathsf{K}}
\Big[
1+\lambda
\Big( 1-\frac{c^2}{\lambda^2}
\sum_{{\mathsf{J}}=1}^{\tilde{\mathsf{K}}-1}
\Big(
\frac{w_{{\mathsf{J}}}}{\kappa_{{\mathsf{J}}}^{N_3}}\Big)^2 
\Big)^{\frac{1}{2}} 
\Big]
$$
has a zero. As $f_{N_3}(\cdot)$ is continuous, it suffices to demonstrate that it attains both negative and positive values. It is obvious that $f_{N_3}(0)<0$. Setting
$$
c_{\max}(N_3)
\;=\;
\lambda \Big(\sum_{{\mathsf{J}}=1}^{\tilde{\mathsf{K}}-1}\Big(\frac{w_{{\mathsf{J}}}}{\kappa_{{\mathsf{J}}}^{N_3}}\Big)^2\Big)^{-1/2}
$$
one observes that 
$$
f_{N_3}(c_{\max}(N_3))
\;=\;
\kappa^{N_3}_{\mathsf{K}}
\Big[
\lambda\left\|(w_{\mathsf{K}},\dots, w_{\tilde{\mathsf{K}}})^{\intercal}\right\|
\Big(
\sum_{{\mathsf{J}}=1}^{\tilde{\mathsf{K}}-1}
\Big(w_{\mathsf{J}}\;\frac{\kappa_{\mathsf{K}}^{N_3}}{\kappa_{{\mathsf{J}}}^{N_3}}\Big)^2
\Big)^{-1/2} -1
\Big]\;.
$$
Since $\frac{\kappa_{\mathsf{K}}}{\kappa_{{\mathsf{J}}}}<1$ for ${\mathsf{J}}\in\{1,\dots,\tilde{\mathsf{K}}-1\}$, positive values are reached for sufficiently large $N_3$.
\hfill $\Box$

\section{Details of the proof of Theorem~\ref{main_result}}

This section contains the proofs of the preparatory lemmas for the proof of Theorem~\ref{main_result}.

\vspace{.2cm}

\noindent {\bf Proof} of Lemma~\ref{deterministic_estimate}.
Inequality~\eqref{deterministic_estimate_inequality} is obviously satisfied if $\mathfrak{a}(v)=0$, as in this case $\mathfrak{a}(\mathcal{R}\cdot v)=0$ holds. Now, let $\mathfrak{a}(v)\neq 0$. Then, its validity is demonstrated by the estimate
\begin{align*}
\begin{split}
\left\|\mathfrak{a}\left(\mathcal{R}\cdot v\right)\right\|^2 & = \left(1+\frac{\left\|\mathfrak{b}\left(\mathcal{R}v\right)\right\|^2+\left\|\mathfrak{c}\left(\mathcal{R}v\right)\right\|^2}{\left\|\mathfrak{a}\left(\mathcal{R}v\right)\right\|^2}\right)^{-1}\\
& \leq \left(1+\frac{\left(\kappa_{\mathsf{L}_{\mathfrak{b}}+\mathsf{L}_{\mathfrak{c}}+1}\right)^2\left\|\mathfrak{b}(v)\right\|^2+\left(\kappa_{\mathsf{L}_{\mathfrak{c}}}\right)^2\left\|\mathfrak{c}(v)\right\|^2}{\left(\kappa_{\mathsf{L}_{\mathfrak{b}}+\mathsf{L}_{\mathfrak{c}}+1}\right)^2\left\|\mathfrak{a}(v)\right\|^2}\right)^{-1}\\
& = \left\|\mathfrak{a}(v)\right\|^2\left(1+\left\|\mathfrak{c}(v)\right\|^2\left[\left(\frac{\kappa_{\mathsf{L}_{\mathfrak{c}}}}{\kappa_{\mathsf{L}_{\mathfrak{b}}+\mathsf{L}_{\mathfrak{c}}+1}}\right)^2-1\right]\right)^{-1}\\
& \leq \left\|\mathfrak{a}(v)\right\|^2\left(1+\left\|\mathfrak{c}(v)\right\|^2\min\left\{1,\left(\frac{\kappa_{\mathsf{L}_{\mathfrak{c}}}}{\kappa_{\mathsf{L}_{\mathfrak{b}}+\mathsf{L}_{\mathfrak{c}}+1}}\right)^2-1\right\}\right)^{-1}\\
& \leq \left\|\mathfrak{a}(v)\right\|^2\left(1-\frac{\left\|\mathfrak{c}(v)\right\|^2}{2}\min\left\{1,\left(\frac{\kappa_{\mathsf{L}_{\mathfrak{c}}}}{\kappa_{\mathsf{L}_{\mathfrak{b}}+\mathsf{L}_{\mathfrak{c}}+1}}\right)^2-1\right\}\right)\,,
\end{split}
\end{align*} 
in which we used that $\left\|\mathfrak{a}(v)\right\|^2+\left\|\mathfrak{b}(v)\right\|^2+\left\|\mathfrak{c}(v)\right\|^2=1$ in the third step. Due to the definition~\eqref{eq-GammaDef} this implies the result.
\hfill $\Box$

\vspace{.2cm}

\noindent {\bf Proof} of Lemma~\ref{bound_average_drift}. Let $\mathscr{U}\in\mbox{\rm O}(\mathsf{L}+1)$ and $s\in [0,1]$. 
We apply the bound $(1+x)^{-1}\leq 1-x+2x^2$ for $x\geq-\frac{1}{2}$ to $x=2\lambda s\left\langle v,Uv\right\rangle +\lambda^2s^2$ where $\lambda\leq \frac{1}{4}$.  This yields the estimate
\begin{align*}
\left\|\left(\mathbf{1}+\lambda s\mathscr{U}\right)v\right\|^{-2}&=\left(1+2\lambda s\left\langle v, \mathscr{U}v\right\rangle+\lambda^2s^2\right)^{-1}\\
&\leq 1-\left(2\lambda s\left\langle v, \mathscr{U}v\right\rangle+\lambda^2s^2\right)+2\left(2\lambda s\left\langle v, \mathscr{U}v\right\rangle+\lambda^2s^2\right)^2\\
&=\left[1-\lambda^2s^2+2\lambda^4s^4\right]-2\lambda s(1-4\lambda^2s^2)\left\langle v,\mathscr{U}v\right\rangle +8\lambda^2s^2\left\langle v,\mathscr{U}v\right\rangle^2
\;.
\end{align*}
As any term of odd order in the entries of $U$ is centered, this implies for the average over $U$
\begin{align*}
\mathbb{E}\,\|\mathfrak{a}((\mathbf{1}+ \lambda sU)\cdot v)\|^2
&\;=\;
\mathbb{E}\left\|\mathfrak{a}\left(\left(\mathbf{1}+\lambda sU\right)v\right)\right\|^2\left\|\left(\mathbf{1}+\lambda sU\right)v\right\|^{-2}\\
&
\;\leq\; \left[1-\lambda^2s^2+2\lambda^4s^4\right]\left(\left\|\mathfrak{a}(v)\right\|^2+\lambda^2s^2\,\mathbb{E}\left\|\mathfrak{a}(Uv)\right\|^2\right)\\
&\qquad -4\,\lambda^2s^2(1-4\lambda^2s^2)\,\mathbb{E}\left\langle \mathfrak{a}(Uv),\mathfrak{a}(v)\right\rangle\left\langle v,Uv\right\rangle\\
&\qquad +\,8\lambda^2s^2\left(\left\|\mathfrak{a}(v)\right\|^2\mathbb{E}\left\langle v,Uv\right\rangle^2+\lambda^2s^2\,\mathbb{E}\left\|\mathfrak{a}(Uv)\right\|^2\left\langle v,Uv\right\rangle^2\right)
\;.
\end{align*}
The averages on the r.h.s. can now be evaluated explicitly, {\it e.g.}, using Lemma~2 in~\cite{LSS},
\begin{align*}
&
\mathbb{E}\left\|\mathfrak{a}(Uv)\right\|^2
\;=\;
\mathbb{E}\textnormal{ tr}\left[U^*\begin{pmatrix}\mathbf{1}_{\mathsf{L}_{\mathfrak{a}}} & 0 
\\ 0 &  0\end{pmatrix} U |v\rangle\langle v|\right]
\;=\;
\frac{\mathsf{L_{\mathfrak{a}}}}{\mathsf{L}+1}
\;,
\\
& \mathbb{E}\left\langle v,Uv\right\rangle^2
\;=\;
\mathbb{E}\textnormal{ tr}\left[U^*|v\rangle\langle v| U |v\rangle\langle v|\right]
\;=\;
\frac{1}{\mathsf{L}+1}\,,
\\
& 
\mathbb{E}\left\|\mathfrak{a}(Uv)\right\|^2\left\langle v,Uv\right\rangle^2
\;=\;
\mathbb{E}\textnormal{ tr}\left[U^*\begin{pmatrix}\mathbf{1}_{\mathsf{L}_{\mathfrak{a}}} &  0 \\ 0  & 0\end{pmatrix}U|v\rangle\langle v| U^* |v\rangle\langle v|U|v\rangle\langle v|\right]
\;=\;
\frac{\mathsf{L}_{\mathfrak{a}}+2\left\|\mathfrak{a}(v)\right\|^2}{\left(\mathsf{L}+1\right)\left(\mathsf{L}+3\right)}\,,
\\
&
\mathbb{E}\left\langle \mathfrak{a}(Uv),\mathfrak{a}(v)\right\rangle\left\langle v,Uv\right\rangle
\;=\;
\mathbb{E}\textnormal{ tr}\left[U^*\begin{pmatrix}\mathbf{1}_{\mathsf{L}_{\mathfrak{a}}} & 0 \\ 0 & 0\end{pmatrix}|v\rangle\langle v| U |v\rangle\langle v|\right]
\;=\;
\frac{\left\|\mathfrak{a}(v)\right\|^2}{\mathsf{L}+1}\;.
\end{align*}
We obtain
\begin{align*}
\mathbb{E}\,\|\mathfrak{a}((\mathbf{1}+ \lambda sU)\cdot v)\|^2
&
\;\leq\; 
\left[1-\lambda^2s^2+2\lambda^4s^4\right]\left(\left\|\mathfrak{a}(v)\right\|^2+\lambda^2s^2\,\frac{\mathsf{L}_{\mathfrak{a}}}{\mathsf{L}+1}\right)\\
&\;\;\;\quad -4\lambda^2s^2(1-4\lambda^2s^2)\,\frac{\left\|\mathfrak{a}(v)\right\|^2}{\mathsf{L}+1}\\
&\;\;\;\quad +8\lambda^2s^2\left(\,\frac{\left\|\mathfrak{a}(v)\right\|^2}{\mathsf{L}+1}+\lambda^2s^2\,\frac{\mathsf{L}_{\mathfrak{a}}+2\left\|\mathfrak{a}(v)\right\|^2}{\left(\mathsf{L}+1\right)\left(\mathsf{L}+3\right)}\right)\\
&
\;=\;
\left(1-\lambda^2s^2\,\frac{\mathsf{L}-3}{\mathsf{L}+1}+\lambda^4s^4\,\frac{16(\mathsf{L}+4)}{(\mathsf{L}+1)(\mathsf{L}+3)}\right)\left\|\mathfrak{a}(v)\right\|^2\\
&\;\;\;\quad +\left(1-\lambda^2s^2\,\frac{\mathsf{L}-5}{\mathsf{L}+3}+2\lambda^4s^4\right)\lambda^2s^2\,\frac{\mathsf{L}_{\mathfrak{a}}}{\mathsf{L}+1}\,.
\end{align*}
This, in turn, implies \eqref{bound_average_drift_inequality}, since $\lambda\leq \frac{1}{4}$ and $\mathsf{L}\geq 3$.
\hfill $\Box$

\vspace{.2cm}

\noindent {\bf Proof} of Lemma~\ref{get_rid_of_unlikely_cases}. Using conditional expectations, one obtains the estimate

\begin{align*}
\mathbb{E}\left\|\mathfrak{a}(v)\right\|^2\left\|\mathfrak{c}(v)\right\|^2
&
\;\geq\; \mathbb{E}\mbox{\small $\left(\left\|\mathfrak{a}(v)\right\|^2\left\|\mathfrak{c}(v)\right\|^2\textnormal{ }\big|\textnormal{ }\left\|\mathfrak{c}(v)\right\|^2\geq \delta\right)$}\textnormal{ }\mathbb{P}\mbox{\small $\left(\left\|\mathfrak{c}(v)\right\|^2\geq\delta\right)$}
\\
&
\;\geq\;
\delta\textnormal{ }\mathbb{E}\mbox{\small $\left(\left\|\mathfrak{a}(v)\right\|^2\textnormal{ }\big|\textnormal{ }\left\|\mathfrak{c}(v)\right\|^2\geq \delta\right)$}\textnormal{ }\mathbb{P}\mbox{\small $\left(\left\|\mathfrak{c}(v)\right\|^2\geq\delta\right)$}
\\
&
\;=\;
\delta\left[\mathbb{E}\left\|\mathfrak{a}(v)\right\|^2-\mathbb{E}\mbox{\small $\left(\left\|\mathfrak{a}(v)\right\|^2\textnormal{ }\big|\textnormal{ }\left\|\mathfrak{c}(v)\right\|^2< \delta\right)$}\textnormal{ }\mathbb{P}\mbox{\small $\left(\left\|\mathfrak{c}(v)\right\|^2<\delta\right)$}\right]
\\
&
\;\geq \;
\delta\left[\mathbb{E}\left\|\mathfrak{a}(v)\right\|^2-\mathbb{P}\mbox{\small $\left(\left\|\mathfrak{c}(v)\right\|^2<\delta\right)$}\right]
\;.
\end{align*} 
This proves \eqref{get_rid_of_unlikely_cases_inequality}.
\hfill $\Box$

\vspace{.2cm}

\noindent {\bf Proof} of Proposition~\ref{prop-StochOd}. The proof is split into two intermediate steps. The first one is similar to Lemma~\ref{lem-2}:

\vspace{.1cm}

\noindent {\bf Step 1.} \textit{Let $v,w\in\mathbb{S}^{\mathsf{L}}$ satisfy $\left\|\mathfrak{c}(v)\right\|=\left\|\mathfrak{c}(w)\right\|$. Then the random variables $\left\|\mathfrak{c}\left(\left(\mathbf{1}+\lambda r U\right)\cdot v\right)\right\|$ and $\left\|\mathfrak{c}\left(\left(\mathbf{1}+\lambda r U\right)\cdot w\right)\right\|$ are distributed identically, that is}
\begin{align}\label{basic_stochastic_order_2}
\mathbb{P}\left(\left\|\mathfrak{c}\left(\left(\mathbf{1}+\lambda r U\right)\cdot v\right)\right\|\in\cdot\right)
\;=\;
\mathbb{P}\left(\left\|\mathfrak{c}\left(\left(\mathbf{1}+\lambda r U\right)\cdot w\right)\right\|\in\cdot\right)
\;.
\end{align}

\vspace{.1cm}

\noindent For the proof, let us first note that  the assumption of $\left\|\mathfrak{c}(v)\right\|=\left\|\mathfrak{c}(w)\right\|$ guarantees the existence of $\left(\mathscr{W}_1,\mathscr{W}_2\right)\in\textnormal{O}(\mathsf{L}_{\mathfrak{a}}+\mathsf{L}_{\mathfrak{b}})\times\textnormal{O}(\mathsf{L}_{\mathfrak{c}})$ such that $\mathscr{W}=\mathscr{W}_1\oplus\mathscr{W}_2\in\textnormal{O}(\mathsf{L}_{\mathfrak{a}}+\mathsf{L}_{\mathfrak{b}})\oplus\textnormal{O}(\mathsf{L}_{\mathfrak{c}})$ satisfies $w=\mathscr{W}v$.
Next let $\tilde{\mathsf{L}}$ be either equal to $\mathsf{L}_{\mathfrak{c}}$ or equal to $\mathsf{L}+1$. Furthermore let $\mathscr{P}_{\tilde{\mathsf{L}}}$ denote the orthogonal projection onto $\{0\}^{\mathsf{L}+1-\tilde{\mathsf{L}}}\times \mathbb{R}^{\tilde{\mathsf{L}}}$. It is obvious that $\mathscr{W}^*$ commutes with $\mathscr{P}_{\tilde{\mathsf{L}}}$. Hence, all $(s,\mathscr{U})\in[0,1]\times\textnormal{O}(\mathsf{L}+1)$ obey
\begin{align*}
\left\|\mathscr{P}_{\tilde{\mathsf{L}}}(\mathbf{1}+\lambda s\mathscr{U})w\right\|^2 
\;=\;
\left\|\mathscr{W}^*\mathscr{P}_{\tilde{\mathsf{L}}}(\mathbf{1}+\lambda s\mathscr{U})\mathscr{W}v\right\|^2
\;=\;
\left\|\mathscr{P}_{\tilde{\mathsf{L}}}(\mathbf{1}+\lambda s\mathscr{W}^*\mathscr{U}\mathscr{W})v\right\|^2
\;.
\end{align*}
This identity is now used in the third equality in the following calculation:
\begin{align*}
\left\|\mathfrak{c}\left((\mathbf{1}+\lambda s\mathscr{U})\cdot w\right)\right\|
&
\;=\;
\left\|\mathfrak{c}\left((\mathbf{1}+\lambda s\mathscr{U})w\right)\right\|\;\left\|(\mathbf{1}+\lambda s\mathscr{U})w\right\|^{-1}
\\
&
\;=\;
\left\|\mathscr{P}_{\mathsf{L}_{\mathfrak{c}}}(\mathbf{1}+\lambda s\mathscr{U})w\right\|\;\left\|\mathscr{P}_{\mathsf{L}+1}(\mathbf{1}+\lambda s\mathscr{U})w\right\|^{-1}
\\
&
\;=\;
\left\|\mathscr{P}_{\mathsf{L}_{\mathfrak{c}}}(\mathbf{1}+\lambda s\mathscr{W}^*\mathscr{U}\mathscr{W})v\right\|\;\left\|\mathscr{P}_{\mathsf{L}+1}(\mathbf{1}+\lambda s\mathscr{W}^*\mathscr{U}\mathscr{W})v\right\|^{-1}
\\
&
\;=\;
\left\|\mathfrak{c}\left((\mathbf{1}+\lambda s\mathscr{W}^*\mathscr{U}\mathscr{W})v\right)\right\|\;\left\|(\mathbf{1}+\lambda s\mathscr{W}^*\mathscr{U}\mathscr{W})v\right\|^{-1}
\\
&
\;=\;
\left\|\mathfrak{c}\left((\mathbf{1}+\lambda s\mathscr{W}^*\mathscr{U}\mathscr{W})\cdot v\right)\right\|\,.
\end{align*}
But $\mathscr{W}^*U\mathscr{W}$ is distributed identically to $U$ so that \eqref{basic_stochastic_order_2} and thus Step 1 follows.

\vspace{.1cm}

In view of Step 1,~\eqref{basic_stochastic_order_inequality} is equivalent to the existence of a path $\phi:[0,1]\rightarrow\mathbb{S}^{\mathsf{L}}$ such that $\left\|\mathfrak{c}(\cdot)\right\|\,\circ\,\phi:[0,1]\rightarrow [0,1]$ is non-decreasing and surjective and that for all $\epsilon\in [0,1]$ the map $t\mapsto \mathbb{P}\big(\left\|\mathfrak{c}\left((\mathbf{1}+\lambda rU)\cdot \phi(t)\right)\right\|\leq \epsilon\big)$ is non-increasing. Hence the proof of the lemma is completed by the following

\vspace{.1cm}

\noindent {{\bf Step 2.}} \textit{The map $f_{\epsilon}:[0,\frac{\pi}{2}]\mapsto [0,1]$ defined by
\begin{align}
\label{basic_stochastic_order_5}
f_{\epsilon}(t)
\;=\;
\mathbb{P}\big(\left\|\mathfrak{c}\left((\mathbf{1}+\lambda rU)\cdot (\cos(t),0,\dots,0,\sin(t))^{\intercal}\right)\right\|\leq\epsilon\big)
\end{align}
is non-increasing for all $\epsilon\in[0,1]$.}

\vspace{.1cm}

\noindent To prove this monotonicity property, it is not necessary to calculate the probability explicitly, but only proportionality is needed. As mentioned in the proof of Theorem~\ref{ac_distribution}, the normalized surface measure $\nu_{\mathsf{L}}$ is distributed identically to the pushforward $\left(h_v\right)_*(\theta_{\mathsf{L}})$ of the Haar measure $\theta_{\mathsf{L}}$ on $\textnormal{O}(\mathsf{L}+1)$ under the map $h_v:\textnormal{O}(\mathsf{L}+1)\rightarrow\mathbb{S}^{\mathsf{L}}$ given by $h_v( U)= Uv$ for any $v\in\mathbb{S}^{\mathsf{L}}$. Thus, $(\mathbf{1}+\lambda rU)\cdot (\cos(t),0,\dots,0,\sin(t))^{\intercal}$ is distributed identically to
\begin{align*}
\frac{(\cos(t)+\lambda rz_1,\lambda rz_2,\dots,\lambda rz_{\mathsf{L}},\sin(t)+\lambda rz_{\mathsf{L}+1})^{\intercal}}{\left\|(\cos(t)+\lambda rz_1,\lambda rz_2,\dots,\lambda rz_{\mathsf{L}},\sin(t)+\lambda rz_{\mathsf{L}+1})^{\intercal}\right\|}
\;,
\end{align*}
where $(z_1,\dots, z_{\mathsf{L}+1})^{\intercal}$ is assumed to be distributed according to $\nu_{\mathsf{L}}$. It follows that
$$
\mathbb{P}\big(\left\|\mathfrak{c}\left((\mathsf{1}+\lambda rU)\cdot (\cos(t),0,\dots,0,\sin(t))^{\intercal}\right)\right\|\leq\epsilon\big)
\;=\;
\mathbb{P}\left(W_{\epsilon}^t(r,z_1,z_{\mathsf{L}_{\mathfrak{a}}+\mathsf{L}_{\mathfrak{b}}+1},\dots, z_{\mathsf{L}+1})\leq 0\right)
\;,
$$
where
\begin{align*}
W_{\epsilon}^t(r,z_1,z_{\mathsf{L}_{\mathfrak{a}}+\mathsf{L}_{\mathfrak{b}}+1},\dots, z_{\mathsf{L}+1})
\;=\;
&
\lambda^2r^2z_{\mathsf{L}+1}^2+2\lambda r (1-\epsilon^2)\sin(t)z_{\mathsf{L}+1}\,+\,\sin^2(t)
\\
& +\lambda^2r^2\|(z_{\mathsf{L}_{\mathfrak{a}}+\mathsf{L}_{\mathfrak{b}}+1},\dots, z_{\mathsf{L}})^{\intercal}\|^2-\left(1+\lambda^2r^2+2\lambda r\cos(t) z_1\right)\epsilon^2
\;.
\end{align*}
Now $W_{\epsilon}^t(r,z_1,z_{\mathsf{L}_{\mathfrak{a}}+\mathsf{L}_{\mathfrak{b}}+1},\dots, z_{\mathsf{L}+1})$ is a parabola in $z_{\mathsf{L}+1}$ with unique minimum. It attains non-positive values if and only if 
\begin{align}
\label{basic_stochastic_order_8}
\left[1+\lambda^2r^2+2\lambda r\cos(t)z_1-(2-\epsilon^2)\sin^2(t)\right]\epsilon^2
\;\geq\; 
\lambda^2r^2\|(z_{\mathsf{L}_{\mathfrak{a}}+\mathsf{L}_{\mathfrak{b}}+1},\dots, z_{\mathsf{L}})^{\intercal}\|^2
\;.
\end{align}
Let us use the notation $n_{\mathfrak{c}}(z)=\|(z_{\mathsf{L}_{\mathfrak{a}}+\mathsf{L}_{\mathfrak{b}}+1},\dots, z_{\mathsf{L}})^{\intercal}\|^2$.
If~\eqref{basic_stochastic_order_8} holds, then the inequality $W_{\epsilon}^t(r,z_1,z_{\mathsf{L}_{\mathfrak{a}}+\mathsf{L}_{\mathfrak{b}}+1},\dots, z_{\mathsf{L}+1})\leq 0$ is equivalent to
\begin{align}\label{basic_stochastic_order_9}
a_{\epsilon,-}^t(r,z_1,n_{\mathfrak{c}}(z))
\;\leq \;
\lambda rz_{\mathsf{L}+1}
\;\leq\;
a_{\epsilon,+}^{t}(r,z_1,n_{\mathfrak{c}}(z))\;,
\end{align}
where the two roots of the polynomial are
$$
a_{\epsilon,\pm}^t(r,z_1,n_{\mathfrak{c}}(z))
\,=\,(\epsilon^2-1)\sin(t)
\pm
\Big[(1+\lambda^2r^2+2\lambda r\cos(t)z_1-(2-\epsilon^2)\sin^2(t))\epsilon^2- \lambda^2r^2n_{\mathfrak{c}}(z)\Big]^{\frac{1}{2}}
.
$$
For later use, let us note that $a_{\epsilon,+}^t(r,z_1,n_{\mathfrak{c}}(z))$ is non-increasing in $t$.

\vspace{.1cm}

Next let $s,\tilde{u}\in[0,1]$ and $u\in [-1,1]$ and set $\rho_{s,u,\tilde{u}}=\lambda s\sqrt{1-u^2-\tilde{u}}$. Now $r$ and $(z_1,z_{\mathsf{L}_{\mathfrak{a}}+\mathsf{L}_{\mathfrak{b}}+1},\dots, z_{\mathsf{L}+1})^{\intercal}$ are independent, and, provided that $\|(z_1,z_{\mathsf{L}_{\mathfrak{a}}+\mathsf{L}_{\mathfrak{b}}+1},\dots, z_{\mathsf{L}})^{\intercal}\|^2=u^2+\tilde{u}$ is fixed, $z_{\mathsf{L}+1}$ is distributed equally to one component of a uniformly distributed vector on the sphere $\sqrt{1-u^2-\tilde{u}}\;\mathbb{S}^{\mathsf{L}_{\mathfrak{a}}+\mathsf{L}_{\mathfrak{b}}-1}$ of radius $\sqrt{1-u^2-\tilde{u}}$. Therefore one has the proportionality for the derivative of the conditional distribution
$$
\frac{\textnormal{d}}{\textnormal{d}x}\mathbb{P}\Big(\lambda rz_{\mathsf{L}+1}\leq x
\,\Big|\,
(r,z_1,n_{\mathfrak{c}}(z))=(s,u,\tilde{u})\Big)
\,\propto\,
\big(1-u^2-\tilde{u}-\tfrac{x^2}{\lambda^2 s^2}\big)^{\frac{\mathsf{L}_{\mathfrak{a}}+\mathsf{L}_{\mathfrak{b}}-3}{2}}
\chi_{[-\rho_{s,u,\tilde{u}},\rho_{s,u,\tilde{u}}]}(x)
\;.
$$
This is similar to \eqref{ac_distribution_0} for $v=e_{\mathsf{L}+1}$. Combining this proportionality relation with \eqref{basic_stochastic_order_9}, still under the assumption that \eqref{basic_stochastic_order_8} is satisfied, one deduces that
\begin{align}
\mathbb{P}\Big( 
W_{\epsilon}^t & (r,z_1,z_{\mathsf{L}_{\mathfrak{a}}+  \mathsf{L}_{\mathfrak{b}}+1},\dots, z_{\mathsf{L}+1})\leq 0
\,\Big|\,
(r,z_1,n_{\mathfrak{c}}(z))=(s,u,\tilde{u})\Big)
\nonumber
\\
& 
\,\propto\,
\int_{a_{\epsilon,-}^{t}(s,u,{\tilde{u}})}^{a_{\epsilon,+}^{t}(s,u,{\tilde{u}})}
\textnormal{d\textbf{x}}(x)\;\big(1-u^2-\tilde{u}-\tfrac{x^2}{\lambda^2 s^2}\big)^{\frac{\mathsf{L}_{\mathfrak{a}}+\mathsf{L}_{\mathfrak{b}}-3}{2}}
\chi_{[-\rho_{s,u,\tilde{u}},\rho_{s,u,\tilde{u}}]}(x)
\;.
\label{basic_stochastic_order_11}
\end{align}
On the other hand, the probability on the l.h.s. vanishes if~\eqref{basic_stochastic_order_8} is violated. By using that the l.h.s. of~\eqref{basic_stochastic_order_8} is non-increasing in $t$ as long as it is non-negative, this is equivalent to $t\in [b_{\epsilon}(r,z_1,z_{\mathsf{L}_{\mathfrak{a}}+\mathsf{L}_{\mathfrak{b}}+1},\dots,z_{\mathsf{L}}),\frac{\pi}{2}]$ for some $b_{\epsilon}(r,z_1,z_{\mathsf{L}_{\mathfrak{a}}+\mathsf{L}_{\mathfrak{b}}+1},\dots,z_{\mathsf{L}})\in [0,\frac{\pi}{2}]$. Hence it suffices to demonstrate that the r.h.s. of \eqref{basic_stochastic_order_11} is non-increasing in $t$ under the condition that \eqref{basic_stochastic_order_8} holds. As one has the inequality $\big|a_{\epsilon,+}^{t}\big(s,u,{\tilde{u}}\big)\big|\leq -a_{\epsilon,-}^{t}\big(s,u,{\tilde{u}}\big)$, it is sufficient to consider the cases
\begin{align*}
& 
\mbox{\rm (i) }\;\;\;\rho_{s,u,\tilde{u}}
\in\big[0,|a_{\epsilon,+}^{t}(s,u,{\tilde{u}})|\big]\,,
\\
&
\mbox{\rm (ii) }\;\;\rho_{s,u,\tilde{u}}
\in\big(|a_{\epsilon,+}^{t}(s,u,{\tilde{u}}\big)|,-a_{\epsilon,-}^{t}(s,u,\tilde{u})\big]\,,
\\
&
\mbox{\rm (iii) }\;\rho_{s,u,\tilde{u}}
\in\big(-a_{\epsilon,-}^{t}(s,u,{\tilde{u}}),\lambda s\big]
\;.
\end{align*}
In these cases the r.h.s. of \eqref{basic_stochastic_order_11} reads respectively:
\begin{align*}
& 
\mbox{\rm (i) }\;\;\;
\chi_{[0,\infty)}\big(a_{\epsilon,+}^{t}(s,u,{\tilde{u}})\big)
\int^{\rho_{s,u,\tilde{u}}}_{-\rho_{s,u,\tilde{u}}}
\textnormal{d\textbf{x}}(x)\;
\big(1-u^2-\tilde{u}-\tfrac{x^2}{\lambda^2 s^2}\big)^{\frac{\mathsf{L}_{\mathfrak{a}}+\mathsf{L}_{\mathfrak{b}}-3}{2}}
\;,
\\
&
\mbox{\rm (ii) }\;\;
\int^{a_{\epsilon,+}^{t}(s,u,{\tilde{u}})}_{-\rho_{s,u,\tilde{u}}}
\textnormal{d\textbf{x}}(x)\;
\big(1-u^2-\tilde{u}-\tfrac{x^2}{\lambda^2 s^2}\big)^{\frac{\mathsf{L}_{\mathfrak{a}}+\mathsf{L}_{\mathfrak{b}}-3}{2}}
\;,
\\
&
\mbox{\rm (iii) }\;
\int^{a_{\epsilon,+}^{t}(s,u,{\tilde{u}})}_{a_{\epsilon,-}^{t}(s,u,{\tilde{u}})}
\textnormal{d\textbf{x}}(x)\;
\big(1-u^2-\tilde{u}-\tfrac{x^2}{\lambda^2 s^2}\big)^{\frac{\mathsf{L}_{\mathfrak{a}}+\mathsf{L}_{\mathfrak{b}}-3}{2}}
\;.
\end{align*}
Now, still under the condition that \eqref{basic_stochastic_order_8} holds, $a_{\epsilon,+}^{t}(s,u,{\tilde{u}})$ is non-increasing in $t$ and, thus so is~\eqref{basic_stochastic_order_11} in the cases (i) and (ii). Moreover, one has the inequality
\begin{align*}
\frac{\textnormal{d}}{\textnormal{d}t}\,
a_{\epsilon,+}^{t}(s,u,{\tilde{u}})
\;\leq \;
\frac{\textnormal{d}}{\textnormal{d}t}\,
a_{\epsilon,-}^{t}(s,u,{\tilde{u}})
\;\leq\; 0
\;.
\end{align*}
If $\mathsf{L}_{\mathfrak{a}}+\mathsf{L}_{\mathfrak{b}}\geq 3$, the case (iii) is therefore dealt with by
\begin{align*}
\big( 1-u^2-\tilde{u}-\tfrac{a_{\epsilon,+}^{t}(s,u,{\tilde{u}})^2}{\lambda^2 s^2}
\big)^{\frac{\mathsf{L}_{\mathfrak{a}}+\mathsf{L}_{\mathfrak{b}}-3}{2}}
\;\geq\;
\big( 1-u^2-\tilde{u}-\tfrac{a_{\epsilon,-}^{t}(s,u,{\tilde{u}})^2}{\lambda^2 s^2}
\big)^{\frac{\mathsf{L}_{\mathfrak{a}}+\mathsf{L}_{\mathfrak{b}}-3}{2}}
\;.
\end{align*}
In conclusion, \eqref{basic_stochastic_order_11} is non-increasing in $t$ for all $s,\tilde{u}\in[0,1]$ and $u\in [-1,1]$. Due to $\mathsf{L}_{\mathfrak{a}}+\mathsf{L}_{\mathfrak{b}}\geq 3$, this finishes the proof of Step 2 and hence also the propostion.
\hfill $\Box$

\vspace{.2cm}

\noindent {\bf Proof} of Corollary~\ref{bound_by_random_phase_system}.
For $w\in\mathbb{S}^{\mathsf{L}}$, $N\in\mathbb{N}$, $\tilde{N}\in\{1,\dots,N\}$ and $M\geq \tilde{N}+1$, let us consider the stochastic order
\begin{align}
\label{bound_by_random_phase_system_1}
\mathbb{P}
\Big(\big\|\mathfrak{c}\big(\big[\prod_{n=\tilde{N}+1}^{M}(\mathbf{1}+\lambda r_nU_n)\big] \mathcal{R}\cdot w\big)\big\|\in\cdot\Big)
\;
\geq_{\textnormal{st}}
\;
\mathbb{P}
\Big(\big\|\mathfrak{c}\big(\big[\prod_{n=\tilde{N}+1}^{M}(\mathbf{1}+\lambda r_nU_n)\big] \cdot w\big)\big\|\in\cdot\Big)
\;.
\end{align}
For $M=\tilde{N}+1$, it follows from Proposition~\ref{prop-StochOd} and the estimate
\begin{align*}
\left\|\mathfrak{c}(w)\right\|
&
\;=\;
\left[1+(\kappa_{\mathsf{L}_{\mathfrak{c}}})^2\left(\left\|\mathfrak{a}(w)\right\|^2+\left\|\mathfrak{b}(w)\right\|^2\right)(\kappa_{\mathsf{L}_{\mathfrak{c}}})^{-2}\left\|\mathfrak{c}(w)\right\|^{-2}\right]^{-\frac{1}{2}}
\\
&
\;\leq \;
\left[1+\left(\left\|\mathfrak{a}(\mathcal{R}w)\right\|^2+\left\|\mathfrak{b}(\mathcal{R}w)\right\|^2\right)\left\|\mathfrak{c}(\mathcal{R}w)\right\|^{-2}\right]^{-\frac{1}{2}}\\
&
\;=\;
\left\|\mathfrak{c}\left(\mathcal{R}\cdot w\right)\right\|
\;,
\end{align*}
holding true whenever $\left\|\mathfrak{c}(w)\right\|>0$. Next we show by an iterative argument that \eqref{bound_by_random_phase_system_1} also holds for larger $M$. This is based on the general fact that the expectations of any non-decreasing function of a pair of stochastically ordered random variable is ordered (see {\it e.g.}~\cite{Kle}, p.~385). Due to \eqref{basic_stochastic_order_2}, the map $g_{\epsilon}:[0,1]\rightarrow [0,1]$ given by
$$
g_{\epsilon}(x)
\;=\;
\mathbb{P}\big(\|\mathfrak{c}((\mathbf{1}+\lambda r U)\cdot \tilde{w})\|>\epsilon\,\big|\,\|\mathfrak{c}(\tilde{w})\|=x\big)
 $$
 is well-defined for all $\epsilon\in [0,1]$. Moreover, it is non-decreasing by Proposition~\ref{prop-StochOd} and can be extended to a non-decreasing function on $\mathbb{R}$. Thus if \eqref{bound_by_random_phase_system_1} holds for some $M\in\{\tilde{N}+1,\dots,N-1\}$, then all $\epsilon\in [0,1]$ satisfy
$$
\mathbb{E}\,
g_{\epsilon}
\Big(\big\|\mathfrak{c}\big([\prod_{n=\tilde{N}+1}^{M}(\mathbf{1}+\lambda r_nU_n)]  \mathcal{R}\cdot w\big)\big\|\Big)
\;\geq\; 
\mathbb{E}\,
g_{\epsilon}
\Big(\big\|\mathfrak{c}\big(\prod_{n=\tilde{N}+1}^{M}(\mathbf{1}+\lambda r_nU_n) \cdot w\big)\big\|\Big)
\;,
$$
or, equivalently,
$$
\mathbb{P}\Big(\big\|\mathfrak{c}\big([\prod_{n=\tilde{N}+1}^{M+1}(\mathbf{1}+\lambda r_nU_n)]  \mathcal{R}\cdot w\big)\big\|
\leq\epsilon\Big)
\;\leq\; 
\mathbb{P}
\Big(\big\|\mathfrak{c}\big(\prod_{n=\tilde{N}+1}^{M+1}(\mathbf{1}+\lambda r_nU_n) \cdot w\big)\big\|\leq\epsilon\Big)
\;,
$$
namely \eqref{bound_by_random_phase_system_1} remains valid if $M$ is replaced by $M+1$ so that it also holds for $M=N$. As $w\in\mathbb{S}^{\mathsf{L}}$ is arbitrary in the above, one infers that all $(v,\epsilon)\in\mathbb{S}^{\mathsf{L}}\times [0,1]$ obey
\begin{align*}
&
\mathbb{P}\Big(\Big\|
\mathfrak{c}\big(\displaystyle\prod_{\tilde{n}=\tilde{N}+1}^N(\mathbf{1}+\lambda r_{\tilde{n}}U_{\tilde{n}})
\prod_{n=1}^{\tilde{N}}\mathcal{R}(\mathbf{1}+\lambda r_nU_n)\cdot v\big)\Big\|\leq\epsilon\Big)
\\
&
\;=\,
\int_{\mathbb{S}^{\mathsf{L}}}\textnormal{d}\mathbb{P}\mbox{\footnotesize $\left(\left(\mathbf{1}+\lambda r_{\tilde{N}}U_{\tilde{N}}\right)\prod_{n=1}^{\tilde{N}-1}\mathcal{R}\left(\mathbf{1}+\lambda r_nU_n\right)\cdot v\in\cdot\right)$}(w)
\,\mathbb{P}\mbox{\footnotesize $\left(\left\|\textnormal{\normalsize$\mathfrak{c}$}\left(\left(\prod_{n=\tilde{N}+1}^{N}\left(\mathbf{1}+\lambda r_nU_n\right)\right) \mathcal{R}\cdot w\right)\right\|\leq \epsilon\right)$}
\\
&
\;\leq\,
\int_{\mathbb{S}^{\mathsf{L}}}\textnormal{d}\mathbb{P}\mbox{\footnotesize $\left(\left(\mathbf{1}+\lambda r_{\tilde{N}}U_{\tilde{N}}\right)\prod_{n=1}^{\tilde{N}-1}\mathcal{R}\left(\mathbf{1}+\lambda r_nU_n\right)\cdot v\in\cdot\right)$}(w)
\,
\mathbb{P}\mbox{\footnotesize $\left(\left\|\textnormal{\normalsize$\mathfrak{c}$}\left(\prod_{n=\tilde{N}+1}^{N}\left(\mathbf{1}+\lambda r_nU_n\right) \cdot w\right)\right\|\leq \epsilon\right)$}
\\
&
\;=\,
\mathbb{P}\Big(\Big\|
\mathfrak{c}\big(\displaystyle\prod_{\tilde{n}=\tilde{N}}^N(\mathbf{1}+\lambda r_{\tilde{n}}U_{\tilde{n}})
\prod_{n=1}^{\tilde{N}-1}\mathcal{R}(\mathbf{1}+\lambda r_nU_n)\cdot v\big)\Big\|\leq\epsilon\Big)
\,.
\end{align*}
An iterative application of this bound yields \eqref{bound_by_random_phase_system_inequality}.
\hfill $\Box$

\vspace{.2cm}

\noindent {\bf Proof} of Proposition~\ref{prop-RPP}.
Let $h\in\mathtt{L}^{\infty}(\mathbb{S}^{\mathsf{L}})$. Using Tonelli's theorem in the second step  as  well as~\eqref{lem-2a-ac} in the penultimate step, one finds
\begin{align*}
\displaystyle\int_{\mathbb{S}^{\mathsf{L}}}\textnormal{d}\nu_{\mathsf{L}}(v)\;\displaystyle\int_{\mathbb{S}^{\mathsf{L}}}\textnormal{d}\varrho_{r,\lambda,v}^{\textnormal{ac}}(w)\;h(w)
&
\;=\;
\displaystyle\int_{\mathbb{S}^{\mathsf{L}}}\textnormal{d}\nu_{\mathsf{L}}(v)\;\displaystyle\int_{\mathbb{S}^{\mathsf{L}}}\textnormal{d}\nu_{\mathsf{L}}(w)\;\frac{\textnormal{d}\varrho_{r,\lambda,v}^{\textnormal{ac}}}{\textnormal{d}\nu_{\mathsf{L}}}(w)\;h(w)
\\
&
\;=\;
\displaystyle\int_{\mathbb{S}^{\mathsf{L}}}\textnormal{d}\nu_{\mathsf{L}}(w)\;\displaystyle\int_{\mathbb{S}^{\mathsf{L}}}\textnormal{d}\nu_{\mathsf{L}}(v)\;\frac{\textnormal{d}\varrho_{r,\lambda,v}^{\textnormal{ac}}}{\textnormal{d}\nu_{\mathsf{L}}}(w)\;h(w)
\\
&
\;=\;
\displaystyle\int_{\mathbb{S}^{\mathsf{L}}}\textnormal{d}\nu_{\mathsf{L}}(w)\;\;h(w)\;\displaystyle\int_{\mathbb{S}^{\mathsf{L}}}\textnormal{d}\nu_{\mathsf{L}}(v)\;\frac{\textnormal{d}\varrho_{r,\lambda,v}^{\textnormal{ac}}}{\textnormal{d}\nu_{\mathsf{L}}}(w)
\\
&
\;=\;
\displaystyle\int_{\mathbb{S}^{\mathsf{L}}}\textnormal{d}\nu_{\mathsf{L}}(w)\;\;h(w)\;\displaystyle\int_{\mathbb{S}^{\mathsf{L}}}\textnormal{d}\nu_{\mathsf{L}}(v)\;\frac{\textnormal{d}\varrho_{r,\lambda,w}^{\textnormal{ac}}}{\textnormal{d}\nu_{\mathsf{L}}}(v)
\\
&
\;=\;
\varrho_{r,\lambda,v}^{\textnormal{ac}}(\mathbb{S}^{\mathsf{L}})\;\displaystyle\int_{\mathbb{S}^{\mathsf{L}}}\textnormal{d}\nu_{\mathsf{L}}(w)\;\;h(w)
\;.
\end{align*}
Combining this with~\eqref{lem-2a-ppsc} yields
\begin{align*}
\int_{\mathbb{S}^{\mathsf{L}}}\textnormal{d}\nu_{\mathsf{L}}(v)\; & \mathbb{E}\; h\left((\mathbf{1}+\lambda rU)\cdot v\right)
\;=\;
\int_{\mathbb{S}^{\mathsf{L}}}\textnormal{d}\nu_{\mathsf{L}}(v)\;\displaystyle\int_{\mathbb{S}^{\mathsf{L}}}\textnormal{d}\varrho_{r,\lambda,v}(w)\;h(w)
\\
&
\;=\;
\int_{\mathbb{S}^{\mathsf{L}}}\textnormal{d}\nu_{\mathsf{L}}(v)\;\displaystyle\int_{\mathbb{S}^{\mathsf{L}}}\textnormal{d}\varrho_{r,\lambda,v}^{\textnormal{pp}}(w)\;h(w)
\;+\;
\displaystyle\int_{\mathbb{S}^{\mathsf{L}}}\textnormal{d}\nu_{\mathsf{L}}(v)\;\displaystyle\int_{\mathbb{S}^{\mathsf{L}}}\textnormal{d}\varrho_{r,\lambda,v}^{\textnormal{ac}}(w)\;h(w)
\\
&
\;=\;
\varrho_{r,\lambda,v}^{\textnormal{pp}}(\mathbb{S}^{\mathsf{L}})\;\displaystyle\int_{\mathbb{S}^{\mathsf{L}}}\textnormal{d}\nu_{\mathsf{L}}(w)\;\;h(w)
\;+\;
\varrho_{r,\lambda,v}^{\textnormal{ac}}(\mathbb{S}^{\mathsf{L}})\;\displaystyle\int_{\mathbb{S}^{\mathsf{L}}}\textnormal{d}\nu_{\mathsf{L}}(w)\;\;h(w)
\\
&
\;=\;
\varrho_{r,\lambda,v}(\mathbb{S}^{\mathsf{L}})\;\displaystyle\int_{\mathbb{S}^{\mathsf{L}}}\textnormal{d}\nu_{\mathsf{L}}(w)\;\;h(w)
\,.
\end{align*}
Since $\varrho_{r,\lambda,v}$ is normalized, the claim \eqref{RPP_equation} follows.
\hfill $\Box$

\pagebreak

\vspace{.2cm}

\noindent {\bf Proof} of Lemma~\ref{probability_RPP}. The measure $\nu_{\mathsf{L}}$ is given by the normalized restriction of the $\mathsf{L}$-dimensional Hausdorff measure in $\mathbb{R}^{\mathsf{L}+1}$ to the sphere $\mathbb{S}^{\mathsf{L}}$ (see~\textit{e.g.}~\cite{Mat}). Let $\mathsf{L}=\mathsf{L}^++\mathsf{L}^-+1$. Later on, we will choose $\mathsf{L}^-=\mathsf{L}_{\mathfrak{c}}-1$. Then let us decompose $v\in\RM^{\mathsf{L}+1}$ as follows
$$
v\;=\;
r\,\begin{pmatrix}
\cos(\theta) v^+ \\
\sin(\theta) v^-
\end{pmatrix}
\;,
$$
where $r=\|v\|$, $\theta\in[0,\frac{\pi}{2}]$ and $v^\pm\in\SM^{\mathsf{L}^\pm}\subset\RM^{\mathsf{L}^\pm+1}$ are unit vectors which are then described by angles $(\theta^\pm_1,\ldots,\theta^\pm_{\mathsf{L}^\pm})\in[0,2\pi)\times [0,\pi)^{\times \mathsf{L}^\pm-1}$ using the standard spherical coordinates, namely $v^\pm=v^\pm(\theta^\pm_1,\ldots,\theta^\pm_{\mathsf{L}^\pm})$ has the components 
$$
v_1^\pm\;=\;\prod^{\mathsf{L}^\pm}_{n=1}\sin(\theta^\pm_n)\;,
\quad
v_k^\pm\;=\;\cos(\theta^\pm_{k-1}) \prod^{\mathsf{L}^\pm}_{n=k}\sin(\theta^\pm_n)\;,
\qquad
v_{\mathsf{L}^\pm+1}^\pm\;=\;\cos(\theta^\pm_{\mathsf{L}^\pm})
\;.
$$
This provides a bijection from  $\RM^{\mathsf{L}+1}$ to $(0,\infty)\times (0,\frac{\pi}{2})\times (0,2\pi)^{\times 2} \times (0,\pi)^{\times \mathsf{L}^++\mathsf{L}^--2}$, up to sets of zero measure. The Jacobian of the transformation is
$$
J
\;=\;
\det\begin{pmatrix}
\cos(\theta)v^+ & -r\sin(\theta)  v^+ & r\cos(\theta)\partial_{\theta^+}v^+ & 0
\\
\sin(\theta)v^- & r\cos(\theta) v^- &  0 & r\sin(\theta)\partial_{\theta^-}v^-
\end{pmatrix}
\;,
$$
which can be evaluated explicitly
$$
J\;=\;r^{\mathsf{L}}\cos(\theta)^{\mathsf{L}^+}\sin(\theta)^{\mathsf{L}^-}
\Big(\prod_{n=1}^{\mathsf{L}^+}\sin(\theta^+_n)^{n-1}\Big) \Big(\prod_{n=1}^{\mathsf{L}^-}\sin(\theta^-_n)^{n-1}\Big)
\;.
$$
Hence
$$
\nu_{\mathsf{L}}(\{v\in\SM^{\mathsf{L}}\;:\;\sin(\theta)^2
\leq \delta\})
\;=\;
\frac{
\int^{\arcsin(\delta^{\frac{1}{2}})}_0 \textnormal{d\textbf{x}}(\theta)\;\sin(\theta)^{\mathsf{L}^-}\cos(\theta)^{\mathsf{L}^+}
}{
\int^{\frac{\pi}{2}}_0 \textnormal{d\textbf{x}}(\theta)\;\sin(\theta)^{\mathsf{L}^-}\cos(\theta)^{\mathsf{L}^+}
}
\;.
$$
Setting $\mathsf{L}^-=\mathsf{L}_{\mathfrak{c}}-1$, substituting $x=\sin(\theta)^2$ and evaluating the integral in the numerator leads to the identity \eqref{eq-SphereEst}. The generalized binomial coefficient can be bounded as follows:
$$
\frac{\Gamma(\tfrac{\mathsf{L}+1}{2})}{
\Gamma(\tfrac{\mathsf{L}_{\mathfrak{c}}}{2})\Gamma(\tfrac{\mathsf{L}_{\mathfrak{a}}+\mathsf{L}_{\mathfrak{b}}}{2})}
\;\leq\;
\frac{\mathsf{L}_{\mathfrak{c}}}{2}
\Big(
\frac{\mathsf{L}+1}{\mathsf{L}_{\mathfrak{a}}+\mathsf{L}_{\mathfrak{b}}}
\Big)^{\frac{\mathsf{L}_{\mathfrak{a}}+\mathsf{L}_{\mathfrak{b}}}{2}-1}
\Big(
\frac{\mathsf{L}+1}{\mathsf{L}_{\mathfrak{c}}}
\Big)^{\frac{\mathsf{L}_{\mathfrak{c}}}{2}}
\;.
$$
%
%
Furthermore, as $\left(\mathsf{L}_{\mathfrak{a}},\mathsf{L}_{\mathfrak{b}}\right)\neq (1,1)$, the factor $(1-x)^{\frac{\mathsf{L}_{\mathfrak{a}}+\mathsf{L}_{\mathfrak{b}}}{2}-1}$ can be bounded by $(1-\frac{x}{2})$ so that the numerator in \eqref{eq-SphereEst} is bounded by
$$
\int^{\delta}_0
\textnormal{d\textbf{x}}(x)
\;
x^{\frac{\mathsf{L}_{\mathfrak{c}}}{2}-1}(1-x)^{\frac{\mathsf{L}_{\mathfrak{a}}+\mathsf{L}_{\mathfrak{b}}}{2}-1}
\;\leq\;
\delta^{\frac{\mathsf{L}_{\mathfrak{c}}}{2}}\Big[
\frac{2}{\mathsf{L}_{\mathfrak{c}}}\,-\,\frac{\delta}{\mathsf{L}_{\mathfrak{c}}+2}\Big]
\;\leq\;
\frac{2}{\mathsf{L}_{\mathfrak{c}}}\,\delta^{\frac{\mathsf{L}_{\mathfrak{c}}}{2}}
\Big[1\,-\,\frac{\delta}{6}\Big]
\;.
$$
This proves \eqref{probability_RPP_inequality}.
\hfill $\Box$

\vspace{.2cm}

\noindent {\bf Proof} of Corollary~\ref{exploit_RPP}.
Let $\delta\in (0,1)$ and $\epsilon\in (0,\min\{\delta,1-\delta\})$. Clearly
$$
\textnormal{supp}(\mathbf{1}+\lambda rU)
\;=\;
\big\{\mathbf{1}+\lambda s\mathscr{U}\;\big|\;(s,\mathscr{U})\in \textnormal{supp}(r)\times\textnormal{O}(\mathsf{L}+1)\big\}
$$
is both contracting and strongly irreducible (see~\cite{BL}, Part~A, Definition III.~1.3 \&~III.~2.1), since $r\not\equiv 0$ and $\lambda>0$. By Furstenberg's theorem, it follows that there is a unique invariant measure which due to Proposition~\ref{prop-RPP} is given by the Haar measure $\nu_{\mathsf{L}}$ on $\mathbb{S}^{\mathsf{L}}$. Furthermore, by  Theorem~III.4.3 in \cite{BL}, one has for any continuous function $h:\mathbb{S}^{\mathsf{L}}\rightarrow \RM$ that
$$
\lim_{N\rightarrow\infty}
\;\sup_{v\in\mathbb{S}^{\mathsf{L}}}\;
\left|\,
\mathbb{E}\;
h\big(\prod_{n=1}^{N}\left(\mathbf{1}+\lambda r_nU_n\right)\cdot v\big)
\;-\;
\int_{\mathbb{S}^{\mathsf{L}}}\textnormal{d}\nu_{\mathsf{L}}({w})\textnormal{ }h({w})
\;\right|
\;=\;
0\;.
$$
Let us choose
$$
h_{\delta,\epsilon}(v)\;=\; 
\min\left\{1,\epsilon^{-1}\left(\delta+\epsilon-\left\|\mathfrak{c}(v)\right\|^2\right)\right\}
\;
\chi_{\{\|\mathfrak{c}(v)\|^2\leq \delta+\epsilon\}}(v)
\;.
$$
By construction, $h_{\delta,\epsilon}$ is continuous. Thus there exists an $\tilde{N_0}= \tilde{N_0}(\mathsf{L},\mathsf{L}_{\mathfrak{c}},\delta,\epsilon)\in\mathbb{N}$ such that all $N\geq \tilde{N_0}$ and $v\in\mathbb{S}^{\mathsf{L}}$ 
\begin{align}\label{exploit_RPP_1B}
\left|\,
\mathbb{E}\;
h_{\delta,\epsilon}\big(\prod_{n=1}^{N}\left(\mathbf{1}+\lambda r_nU_n\right)\cdot v\big)
\;-\;
\int_{\mathbb{S}^{\mathsf{L}}}\textnormal{d}\nu_{\mathsf{L}}(v)\textnormal{ }h_{\delta,\epsilon}(v)
\;\right|
\;\leq \;
\epsilon\;.
\end{align}
Further, $h_{\delta,\epsilon}$ can be bounded from below and above by indicator functions:
\begin{align}\label{exploit_RPP_2}
\chi_{\{\|\mathfrak{c}(v)\|^2\leq \delta\}}
\;\leq \;
h_{\delta,\epsilon}
\;\leq\;
\chi_{\{\|\mathfrak{c}(v)\|^2\leq \delta+\epsilon\}}
\;.
\end{align}
Now using~\eqref{exploit_RPP_1B}, \eqref{exploit_RPP_2} as well as \eqref{probability_RPP_inequality} with $\delta+\epsilon$ instead of $\delta$ it follows that
\begin{align*}
\mathbb{P}
\Big(\big\|\mathfrak{c}\big(\prod_{n=1}^{N}(\mathbf{1}+\lambda r_nU_n)\cdot v\big)\big\|^2<\delta\Big) 
&
\;\leq\; \mathbb{E}\;h_{\delta,\epsilon}
\Big(\prod_{n=1}^{N}(\mathbf{1}+\lambda r_nU_n)\cdot v\Big)
\\
&
\;\leq\; 
\int_{\mathbb{S}^{\mathsf{L}}}\textnormal{d}\nu_{\mathsf{L}}(v)\textnormal{ }h_{\delta,\epsilon}^{\mathsf{L}}(v)+\epsilon
\\
&
\;\leq\; 
\nu_{\mathsf{L}}\big(\big\{\|\mathfrak{c}(v)\|^2\leq \delta+\epsilon\big\}\big)
\,+\,\epsilon
\\
&
\; \leq \;
\Big(\frac{\mathsf{L}+1}{\mathsf{L}_{\mathfrak{a}}+\mathsf{L}_{\mathfrak{b}}}\Big)^{\frac{\mathsf{L}_{\mathfrak{a}}+\mathsf{L}_{\mathfrak{b}}}{2}-1}
\Big(\frac{\mathsf{L}+1}{\mathsf{L}_{\mathfrak{c}}}\,(\delta+\epsilon)\Big)^{\frac{\mathsf{L}_{\mathfrak{c}}}{2}}
\Big(1-\frac{\delta}{6}\Big)
\,+\,\epsilon
\\
&
\; = \;
\Big(\frac{\mathsf{L}+1}{\mathsf{L}_{\mathfrak{a}}+\mathsf{L}_{\mathfrak{b}}}\Big)^{\frac{\mathsf{L}_{\mathfrak{a}}+\mathsf{L}_{\mathfrak{b}}}{2}-1}
\Big(\frac{\mathsf{L}+1}{\mathsf{L}_{\mathfrak{c}}}\,\delta\Big)^{\frac{\mathsf{L}_{\mathfrak{c}}}{2}}
\;-\;\eta(\epsilon,\delta)
\;,
\end{align*}
the last equation simply by definition of $\eta(\epsilon,\delta)$. Now one readily checks that $\lim_{\epsilon\downarrow 0}\eta(\epsilon,\delta)$ is positive and therefore, by continuity of $\eta(\epsilon,\delta)$, there exists a positive $\epsilon$ for which the bound \eqref{exploit_RPP_inequality} is satisfied for some positive $\eta=\eta(\mathsf{L},\mathsf{L}_{\mathfrak{c}},\delta)>0$.
\hfill $\Box$

\vspace{.2cm}



\begin{thebibliography}{99}

\bibitem{AW} M.~Aizenman, S.~Warzel, {\sl Random operators}, (AMS, Providence, 2015). 

\bibitem{BaR} S.~Bachmann,  W.~De~Roeck, {\sl From the Anderson model on a strip to the DMPK equation and random matrix theory}, J. Stat. Phys. {\bf 139}, 541-564 (2010).

\bibitem{BL} P.~Bougerol, J.~Lacroix, {\sl Products of Random Matrices with Applications to Schr{\"o}dinger Operators}, (Birkh{\"a}user, Boston, 1985).


\bibitem{CL} R.~Carmona, J.~M.~Lacroix, {\sl Spectral theory of random Schr\"odinger operators}, (Birk\-h\"auser, Basel, 1990).




\bibitem{GM} Ya.~I.~Goldsheid, A.~G.~Margulis, {\sl Lyapunov exponents of a product of random matrices},
Russian Math. Surveys {\bf 44}, 11-71 (1989).

\bibitem{Kle} A.~Klenke, {\sl Probability Theory: A Comprehensive Course}, (Springer Universitext, London,  2013).

\bibitem{Kal} O.~Kallenberg, {\sl Foundations of Modern Probability}, (Springer, New York, 2002).


\bibitem{LSS} A.~W.~W.~Ludwig, H.~Schulz-Baldes, M.~Stolz, {\sl Lyapunov Spectra for All Ten Symmetry Classes of Quasi-one-dimensional Disordered Systems of Non-interacting Fermions}, J. Stat. Phys. {\bf 152}, 275-304 (2013).

\bibitem{Mat} P.~Mattila, {\sl Geometry of Sets and Measures in Euclidean Spaces: Fractals and rectifiability}, (Cambridge University Press, Cambridge, 1995).



\bibitem{RS} R.~R\"omer, H.~Schulz-Baldes, {\sl Weak disorder expansion for localization lengths of quasi-1D systems}, Euro. Phys. Lett. {\bf 68}, 247-253 (2004).

\bibitem{RS1} R.~R\"omer, H.~Schulz-Baldes, {\sl Random phase property and the Lyapunov spectrum for disordered multi-channel systems}, J. Stat. Phys. {\bf 140}, 122-153 (2010).

\bibitem{SS} C.~Sadel, H.~Schulz-Baldes, {\sl Random Lie group actions on compact manifolds: a perturbative analysis}, Annals of Probability {\bf 38}, 2224-2257 (2010).

\bibitem{SV} C.~Sadel,  B.~Vir\'ag, {\sl A central limit theorem for products of random matrices and GOE statistics for the Anderson model on long boxes}, Commun. Math. Phys. {\bf 343}, 881-919, (2016).


\bibitem{SB1} H.~Schulz-Baldes, {\sl Perturbation theory for an Anderson model on a strip}, GAFA {\bf 14}, 1089-1117 (2004).

\bibitem{VV} B.~Valk\'o, B.~Vir\'ag, {\sl Random Schr\"odinger operators on long boxes, noise explosion and the GOE}, Trans. AMS {\bf  366}, 3709-3728 (2014).


\end{thebibliography}
\end{document}